\pdfoutput=1
\documentclass[%
 reprint,
nobibnotes,
 amsmath,amssymb,
 aps,
prb,
]{revtex4-2}

\usepackage{graphicx}% Include figure files
\usepackage{dcolumn}% Align table columns on decimal point
\usepackage{bm}% bold math hypertext 
\usepackage[backref=none, breaklinks=True]{hyperref} %% This provides \href command
\usepackage[usenames,dvipsnames]{xcolor}
\usepackage{color}
\usepackage[framemethod=TikZ]{mdframed}
\usepackage{import}

\usepackage{algorithm}
\usepackage{algpseudocode}
\usepackage{multirow}
\usepackage{longtable}
\usepackage{dcolumn}
\usepackage{booktabs}
\newcommand\myshade{85}
\colorlet{mylinkcolor}{violet}
\colorlet{mycitecolor}{YellowOrange}
\colorlet{myurlcolor}{Aquamarine}
\hypersetup{
  linkcolor  = mylinkcolor!\myshade!black,
  citecolor  = mycitecolor!\myshade!black,
  urlcolor   = myurlcolor!\myshade!black,
  colorlinks = true,
}

\newcommand{\ket}[1]{\left| #1 \right \rangle}

\newcommand{\braketmatrix}[3]{\left \langle #1 \middle| #2 \middle| #3 \right \rangle}

\begin{document}

\preprint{APS/123-QED}

\title{Quantum variational rewinding for time series anomaly detection}% Force line breaks with \\

\author{Jack S. Baker$^1$}
\email{jack@agnostiq.ai}

\author{Haim Horowitz$^1$}
% \email{haim@agnostiq.ai}

\author{Santosh Kumar Radha$^1$}%
% \email{santosh@agnostiq.ai}
\affiliation{$^1$Agnostiq Inc., 325 Front St W, Toronto, ON M5V 2Y1}

\author{Stenio Fernandes$^2$}

\author{Colin Jones$^2$}

\author{Noorain Noorani$^2$}

\author{Vladimir Skavysh$^2$}%
\email{skav@bankofcanada.ca}
\affiliation{$^2$Bank of Canada, 234 Wellington Street Ottawa, ON K1A 0G9}

\author{Philippe Lamontagne$^3$}%
\email{philippe.lamontagne2@nrc-cnrc.gc.ca}
\affiliation{$^3$National Research Council Canada, 5145 av.\ Decelles, Montréal, QC, H3T 2B2}

\author{Barry C. Sanders$^4$}%
\email{sandersb@ucalgary.ca}
\affiliation{$^4$Institute for Quantum Science and Technology, University of Calgary, Alberta T2N 1N4, Canada}

\date{\today}% It is always \today, today,

\begin{abstract}
 Electron dynamics, financial markets and nuclear fission reactors, though seemingly unrelated, all produce observable characteristics evolving with time. Within this broad scope, departures from normal temporal behavior range from academically interesting to potentially catastrophic. New algorithms for time series anomaly detection (TAD) are therefore certainly in demand. With the advent of newly accessible quantum processing units (QPUs), exploring a quantum approach to TAD is now relevant and is the topic of this work. Our approach - \textit{Quantum Variational Rewinding}, or, QVR - trains a family of parameterized unitary time-devolution operators to cluster normal time series instances encoded within quantum states. Unseen time series are assigned an anomaly score based upon their distance from the cluster center, which, beyond a given threshold, classifies anomalous behaviour. After a first demonstration with a simple and didactic case, QVR is used to study the real problem of identifying anomalous behavior in cryptocurrency market data. Finally, multivariate time series from the cryptocurrency use case are studied using IBM’s \texttt{Falcon r5.11H} family of superconducting transmon QPUs, where anomaly score errors resulting from hardware noise are shown to be reducible by as much as 20\% using advanced error mitigation techniques.
\end{abstract}

\maketitle

\section*{I. \quad Introduction}

The task of identifying abnormal or \textit{anomalous} behaviour in time series data is known as time series anomaly detection (TAD). Although there exists a taxonomy of problems and variations under the banner of TAD \cite{BlzquezGarca2022}, arguably the most practically relevant is novelty detection. Within machine learning, novelty detection is a one-class classification problem~\cite{Seliya2021} where learning is performed using a set of time series known to exhibit \textit{normal} (i.e, non-anomalous) behaviour~\cite{Gupta2014, BlzquezGarca2022}. Subsequently, anomalies are classified by their deviation from the learnt normal model. The practical relevance of novelty detection stems from the relative scarcity of labeled anomalies to normal data in a broad array of applications spanning fraud detection~\cite{Ferdousi2006}, intrusion detection~\cite{Duque_Anton_2018}, medical diagnosis~\cite{Agliari2020, Homayouni2021}, observational astronomy~\cite{Zhang2018} and many others. This wide application domain paired with the advent of newly accessible quantum processing units \cite{Gyongyosi2019} (QPUs) naturally poses the question: \textit{Can a quantum approach to TAD be devised?}.
\par

In the context of central banking, an emerging application of TAD is analyzing cryptocurrency time series data. Of particular importance are Bitcoin---which is the first and most popular cryptocurrency in the world and which has already been used as legal tender \cite{Perez2021}---and stablecoins (cryptocurrencies partially or fully pegged to safe assets, such as USD)%, such as Tether \cite{tether2022}--the
. This is of interest to central banks because of their role in oversight over payment service providers \cite{retailact} and because of how cryptocurrency developments affect the issuance of central bank digital currency \cite{lane2021}. Despite seeing increased use, cryptocurrencies and stablecoins remain flawed as a method of payment \cite{lane2021}. To monitor risk and possible losses in stablecoins, central banks are developing frameworks to classify stablecoins, identify risk scenarios for various classes of stablecoins, and quantify possible losses stemming from these risk scenarios (see for instance \cite{garcia2021stablecoin}). In this context, it is important for central banks to investigate novel tools for cryptocurrency time series analysis and anomaly detection including quantum algorithmic approaches.

In state-of-the-art classical TAD, it is popular to use different flavors of convolutional neural networks~\cite{Wen2019, Kim2019, Hsu2020} to encode temporal features including long-short-term-memory~\cite{Hochreiter1997, malhotra2015long, shipmon2017, Zhang2018, Ji2021}, bi-directional long-short-term-memory~\cite{Graves2005, Zhang2021} and gated recurrent units ~\cite{Kyunghyun2018, Guo2018}. These techniques are now used in tandem with other network topologies giving rise to complex hybrid networks including recurrent autoencoders~\cite{Pereira2018, Kieu2019} and variational autoencoder-generative adversarial networks~\cite{Zijian2020}. With the advent of deep learning \cite{Chalapthy2019Survey}, these techniques can be stacked as layers of a larger and more complex artificial network to achieve superior performance on large data sets \cite{Shone2018}. \par

In this work, we do not invoke such artificial complexity. Instead, we encode temporal features in the natural time evolution of quantum systems as part of a quantum machine learning (QML)~\cite{Biamonte2017} algorithm suitable for gate-model QPUs. As we shall demonstrate, TAD (with a focus on novelty detection) can be achieved through exploiting the power of unitary time-devolution operators. In short, parameterized unitary operators are trained to time-devolve quantum states encoded with time series data representative of the normal behaviour of a system. Training tunes the parameters of the unitary operators to produce quantum states whose expectation values given a general observable are clustered about a center. Then, given a new and unseen time series as input, the trained model produces expectation values lying some distance from the normal cluster center. Within some distance threshold, this time series is classified as normal. Otherwise, it is anomalous. At odds with data-hungry deep learning, our chosen time-devolution technique \cite{Crstoiu2020} is parameterizable in a way such that it becomes a member of a class of quantum models with low generalization error using a small number of training examples \cite{Caro2022}.

In designing such an algorithm, care should be taken to ensure its suitability for present-generation and near-future QPUs. That is, presently,
we are in the near-intermediate scale quantum (NISQ) era~\cite{Preskill2018} where limited numbers of noisy qubits must be leveraged in low circuit depth applications. So far, the most promising of approaches found to satisfy these constraints are known as \textit{variational quantum algorithms} (VQAs)~\cite{Cerezo2021}:
algorithms whose cost function is (at least in part) defined using parameterized quantum circuits and optimized within a hybrid quantum-classical loop. VQAs suited for static (i.e, without time dependence) anomaly detection have been proposed~\cite{Liu2018, Liang2019, Kotmann2021, Alona2021, Herr2021}, but, none have been designed explicitly to encode temporal features. Naively, such encoding can be achieved by application of Trotterized time-evolution operators \cite{Wiebe2010}. However, in most cases, the resulting quantum circuits are too deep for present NISQ hardware \cite{Crstoiu2020}. The recently proposed \textit{variational fast forwarding} (VFF)~\cite{Crstoiu2020} approach promises to alleviate this problem for certain types of Hamiltonian, replacing Trotterization with a low-depth variational Ans\"{a}tz with known error bounds \cite{Crstoiu2020}. \par
% \bcs{We can just say `emerging' rather than `beginning to emerge'.} - absolutely
% \bcs{I trimmed wording here but saved as a commented-out line. I see lots of places to trim wording, which I can do if others agree and we want to make our paper `crisper'.
% Note: this previous sentence is convoluted and confusing for me, even after making it `crisper'.
% Are we saying that VQAs are immature but VFF is a key part and is currently feasible?} - The point here is that static stuff has be possible previously because they do not deal with the harder task of encoding temporal features. VFF makes this possible at shallow depth. I will try and make this point a little clearer
% \bcs{I am not clear on this preceding sentence.
% The Lie-Trotter expAnsion (why not include Suzuki?)
% is an ordered operator expAnsion.
% Approximate diagonalisation is a different beast.
% I think we should say something about replacing the quantum circuit for one by the quantum circuit by the other and give appropriate references at the end of this sentence.} -  I did this: basically, replace Trotterized circuits for VFF Ans\"{a}tz.

Algorithms exploiting VFF for other purposes have recently been proposed~\cite{Radha2021, Horowitz2022, Gibbs2022, Caro2022}. Applied to combinatorial optimization, XY-mixing operators used for enforcing constraints in the Quantum Alternating Operator Ans\"{a}tz~\cite{Hadfield2019} were learnt at a lower depth than their usual Trotterized implementation~\cite{Radha2021}. A quantum generative model used VFF to model the quantum processes underlying observed time series data~\cite{Horowitz2022}. Having learnt these processes, synthetic time series instances were generated, capturing complex temporal relationships present in the real time series. Lastly, in~\cite{Gibbs2022} and~\cite{Caro2022} the generalization bounds for this class of models were studied and ``resource efficient fast forwarding"~\cite{Gibbs2022} was demonstrated to approximately diagonalize Heisenberg and XY-Hamiltonians. \par

% I like this paragraph a lot but I think it should come later when we either present the formalism or demonstrate parallel quantum execution.
% When implemented within our algorithm, quantum computations at different points in time become independent from one-another and thus are \textit{embarrassingly parallel} tasks;
% % \bcs{If we quote something, we should give the source of the quote.} - very difficult to locate the origin of the phrase Embarrassingly Parallel. It is disputed and known enough that I don't think it needs citation. Rather than quotations, I will just put italics.
% the algorithm can utilize multiple QPUs simultaneously. Paired with the low number of quantum operations per circuit (i.e, low depth), our algorithm is a canonical example of the quantum analogue to the many-task computation paradigm~\cite{raicu2008, raicu2009many}. \par

% \textcolor{cyan}{Jack: Mention exponential speedup of quantum simulation on QPUs: however, this doesn't mean there is advantage for a ML problem...} \par

In what follows, we devise the Quantum Variational Rewinding (QVR) algorithm for TAD, presenting it in a modular way, agnostic to particular forms of circuit Ans\"{a}tze, observables and other components of the algorithm, showing that our approach is deeply customizable with the potential for many algorithmic variations. Then, using particular algorithmic variations, we employ QVR for two examples: (1) noisy univariate data with synthetic anomalies and (2) multivariate cryptocurrency trading data in the presence of large movements of cryptocurrency on the blockchain. Case~(1) is intended as a didactic example whereas Case~(2) is a real and useful application aiming to understand how information from the blockchain affects the behavior of participants in cryptocurrency trading markets. To show that the algorithm is suitable for NISQ era QPUs, for Case~(2), we demonstrate the evaluation of anomaly scores and classification performance on IBM's \texttt{Falcon r5.11H} family of noisy superconducting transmon chips (\texttt{ibm\_lagos}, \texttt{ibm\_perth} and \texttt{ibm\_jakarta}), showing performance enhancement using advanced error mitigation techniques. Although QVR does \textit{not} require high depth circuits, it does require a large number of measurements from low depth circuits. Therefore, to make QVR tractable using presently available QPU resources, parallel QPU execution is paired with high-speed near-QPU classical compute which together reduce execution times to a tolerable level.

\section*{II. \quad Quantum Variational Rewinding}

\begin{figure}
    \centering
    \includegraphics[width=\linewidth]{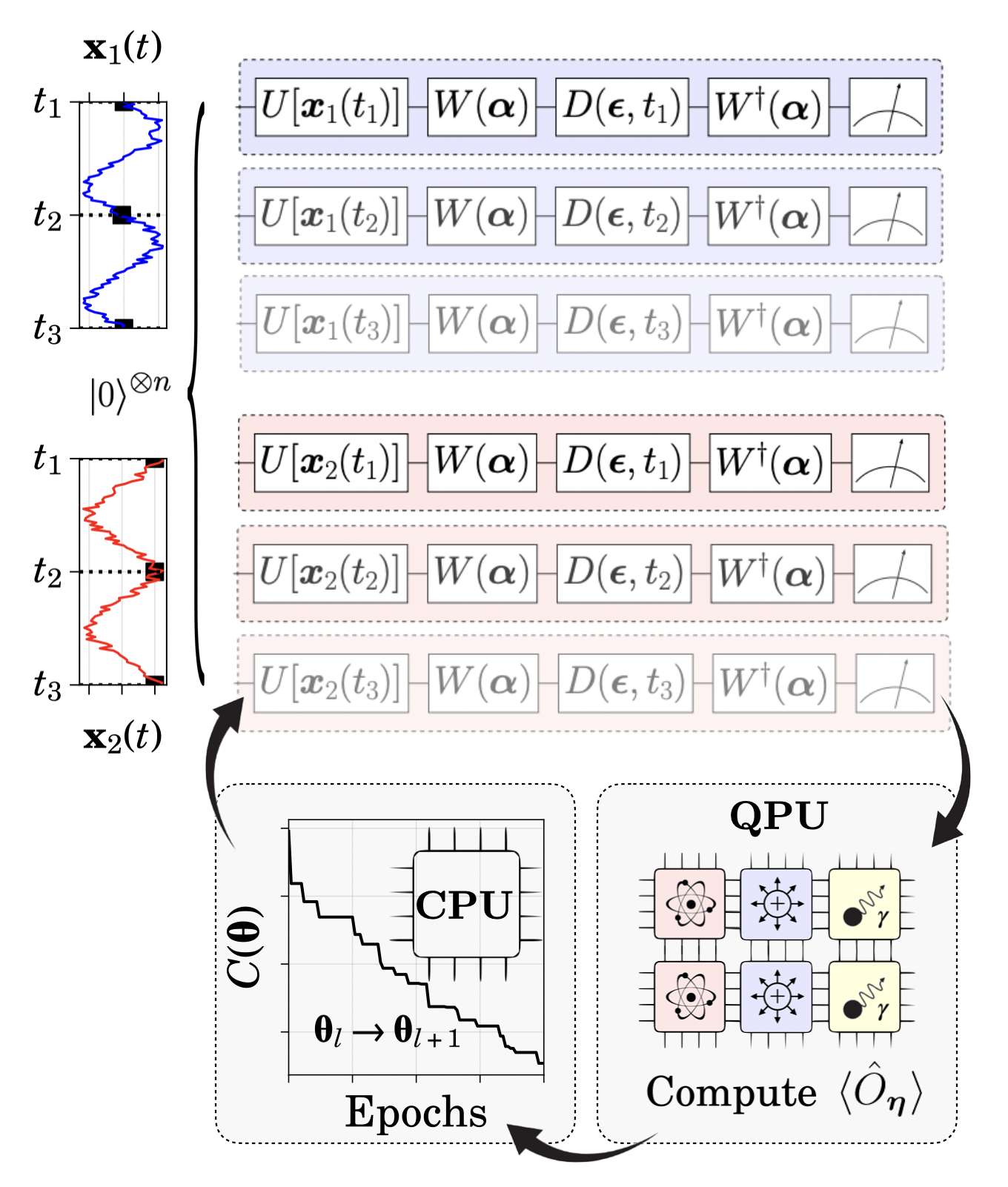}
    \caption{The hybrid quantum-classical loop used for training the QVR model. Starting from the top, $N_X \times N_{\mathcal{T}} \times N_E$ quantum circuits are encoded with time series at selected points in time (shown to the left) followed by the Ans{\"a}tz described in Eq. \ref{eq:vff}. Following the downwards arrow, circuits are offloaded to (multiple) QPUs to prepare $|\bm{x}_i(t_j), \bm{\alpha}, \bm{\epsilon} \rangle$ (Eq. \ref{eq:state}) to evaluate $\langle \hat{O}_{\bm{\eta}} \rangle$ (Eq. \ref{eq:expecatation}) for each. Moving left, all evaluated $\langle \hat{O}_{\bm{\eta}} \rangle$ are transferred to a classical computer and used to build the cost function at the $l^{\text{th}}$ iteration of the quantum-classical loop: $C(\bm{\theta}_l)$ (Eq. \ref{eq:cost}). Using a classical optimization routine, a new candidate $\bm{\theta}_{l+1}$ is found and used to generate the quantum circuits in the next iteration of the loop.}
    \label{fig:hero}
\end{figure}

Now motivated, we present the formalism of QVR and discuss its conceptual underpinning. Throughout this discussion, Fig. \ref{fig:hero} should be used as a companion, providing a graphical interpretation of QVR. Full pseudocode for QVR can be found in the Supplemental Material (SM) \cite{supplement}. In the following, $m$ will denote the size of the training set that we consider, $d$ will be a fixed integer such that each element from the training set is a $d$-dimensional time series, and $p$ will be a fixed integer such that each time series consists of $p$-many time steps. Our ultimate goal will be to learn a family of unitary matrices that, together with the cost function (to be defined later in Eq. \ref{eq:cost}), will give rise to a classification criterion that will allow us to determine whether a given time series is normal or anomalous. \par

We begin with a training set $X = \{ \bm{x}_1(t), \bm{x}_2(t), \ldots, \bm{x}_m(t) \}$ of $d$-dimensional real-valued discrete functions of time (time series) $\bm{x}_i(t) = \{x^1_i(t), x^2_i(t), \ldots, x^d_i(t) \}$ where $t \in \mathcal{T} = \{t_1, t_2, \ldots, t_p \}$. We shall refer to the elements of $\mathcal T$ as time points, which can be general real numbers. Using this training set, our goal is to learn an optimal set of parameters (where the exact sense of ``optimal" is explained in the rest of this Section) that will give rise to a  real-valued function $a_X[\bm{y}]$ that is able to assign anomaly scores to general $d$-dimensional time series $\bm{y}$. By comparing this anomaly score to a set threshold, we determine if $\bm{y}$ is normal or anomalous. \par

Explicitly, we train our model with mini-batches of the number of series $B_X \subseteq X$ and of the number of time points in each series $B_{\mathcal{T}} \subseteq \mathcal{T}$. That is, for each iteration of the quantum-classical loop shown in Fig. \ref{fig:hero}, we draw randomly $N_X=|B_X|$-many time series restricted to randomly chosen $N_{\mathcal{T}} = |B_T|$-many time points. With these mini-batches, we prepare the states $U[\bm{x}_i(t_j)] \ket{0}^{\otimes n} = \ket{\bm{x}_i(t_j)}$ for embedding unitary $U[\bm{x}_i(t_j)]$ acting on the $n$ qubit state of zeroes. This step applies a feature map to the classical data, embedding it as a quantum state \cite{Schuld2020}. The explicit choice of $U[\bm{x}_i(t_j)]$ will be given when a specific algorithmic variation is described.  It should be noted now that while examples in this work deal with quantum states created by a chosen embedding of classical data, we conjecture that our approach is also applicable to quantum data (see \cite{Huang2021}, for example). That is, should one wish to study the temporal behaviour of a set of natively occurring quantum states $\Psi = \{|\psi_1(t) \rangle, |\psi_2(t) \rangle, \ldots, |\psi_m(t) \rangle \}$, one could skip the embedding step and proceed with the native quantum states directly. I.e, the training data set would become $X = \Psi$. In this setting, QVR would then benefit from known exponential advantages in sampling complexity \cite{Aharonov2022, Huang2021}. Testing this approach for quantum data is outside the scope of this work thus we leave evaluating the success of such an approach as an open problem. Classical or quantum data, each state now undergoes a dynamical process (where our notation continues following the classical data case)
\begin{equation}
    \ket{\bm{x}_i(t_j), \bm{\alpha}, \bm{\epsilon}} := e^{-i\hat{H}(\bm{\alpha}, \bm{\epsilon})t_j}  \ket{\bm{x}_i(t_j)}
    \label{eq:state}
\end{equation}
as generated by a general parameterized Hamiltonian operator $\hat{H}(\bm{\alpha}, \bm{\epsilon})$. A family of dynamical processes can be implemented with controllable circuit depth using the structure first proposed in \cite{Crstoiu2020}
\begin{equation}
    e^{-i\hat{H}(\bm{\alpha}, \bm{\epsilon})t_j} = W^{\dagger}(\bm{\alpha}) D(\bm{\epsilon}, t_j) W(\bm{\alpha})
    \label{eq:vff}
\end{equation}
for parameterized unitary 
$W(\bm{\alpha})$, diagonal time-encoded unitary $D(\bm{\epsilon}, t_j)$, variational parameters $\bm{\alpha}$ and $2^n$-length vector $\bm{\epsilon}$. It can then be seen that Eq. \ref{eq:vff} is an eigendecomposition of $e^{-i\hat{H}(\bm{\alpha}, \bm{\epsilon})t_j}$ where $W(\bm{\alpha})$ represents the matrix of eigenvectors and $D(\bm{\epsilon}, t_j)$ its diagonalized form. The unitary operation that corresponds to $D(\bm{\epsilon}, t_j)$ is implemented using a $k$-local approximation of $e^{-iM(\bm{\epsilon})t_j}$ where $M(\bm{\epsilon})$ is an intermediate non-unitary diagonal matrix. Here, $k$-locality is the restriction that every separable term of $D(\bm{\epsilon}, t_j)$ can only be approximated with the mutual interaction of $k$-many qubits. We use the procedure defined in~\cite{Welch2014} meaning that while $D(\bm{\epsilon}, t_j)$ is a diagonal matrix with $2^n$ entries, we can implement a $k$-local version using a $\operatorname{poly}(n)$ number of gates~\cite{Welch2014} and need not store $\sim 2^n$ eigenvalues $\bm{\epsilon}$ in memory. \par

While Eq. \ref{eq:vff} would conventionally be interpreted as a forward time-evolution operator, in the context of our algorithm, we conceptualize Eq. \ref{eq:vff} as a devolving (i.e backwards in time) $\ket{\bm{x}_i(t_j)}$ by time $t_j$. This is permitted since a forward-time evolution as generated $\hat{H}(\bm{\alpha}, \bm{\epsilon})$ is equivalent to a devolution as generated $-\hat{H}(\bm{\alpha}, \bm{\epsilon}) = \hat{H}(\bm{\alpha}, \bm{\epsilon})^{\prime}$ where $\hat{H}(\bm{\alpha}, \bm{\epsilon})^{\prime}$ is another valid Hamiltonian. Also, taking the conjugate transpose of both sides of Eq. \ref{eq:vff}, we find $ e^{i\hat{H}(\bm{\alpha}, \bm{\epsilon})t_j} ={e^{-i\hat{H}(\bm{\alpha}, \bm{\epsilon})t_j}}^{\dagger}= W^{\dagger}(\bm{\alpha}) D(\bm{\epsilon}, -t_j) W(\bm{\alpha})=W^{\dagger}(\bm{\alpha}) D(-\bm{\epsilon}, t_j) W(\bm{\alpha})$. This is an important insight as we will later learn that $\bm{\epsilon}$ is sampled from normal distributions with variable centres which are permitted to be positive or negative. This means that both $D(-\bm{\epsilon}, t_j)$ and $D(\bm{\epsilon}, t_j)$ can be sampled at different model parameters which are determined in training. Also, unlike Trotterized evolution, any time $t$ can be accessed without first traversing through intermediate time steps. In the context of forward time evolution, the authors of Ref. \cite{Crstoiu2020} interpret this as a \textit{fast forwarding} whereas in the context of devolution we interpret it as a \textit{rewinding}. This is the core principle of QVR. \par

% generated by $-\hat{H}(\bm{\alpha}, \bm{\epsilon})$ (in other words, we consider the time reversal $t\mapsto -t$). We therefore refer to Equation \ref{eq:vff} as a \textit{rewinding operator} and is the key ingredient to QVR. \par

When preparing Eq. \ref{eq:state}, $\bm{\epsilon}$ is uniformly randomly drawn from independent normal distributions $\epsilon_q \sim \mathcal{N}(\mu_q, \sigma_q)$, $q \in \{1, 2, \ldots Q| Q \leq 2^n -1 \}$ which we abbreviate to $\bm{\epsilon} \sim \mathcal{N}(\bm{\mu}, \bm{\sigma})$. That is, the vector $\bm{\epsilon}$ from \ref{eq:state} will be randomly generated multiple times in order to compute the expectation value below. For a single random $\bm{\epsilon}$, we evaluate the expectation value of a general parameterized observable $\hat{O}_{\bm{\eta}}$
\begin{equation}
\Omega \left(\bm{x}_i(t_j), \bm{\alpha}, \bm{\epsilon}, \bm{\eta} \right) :=  \braketmatrix{\bm{x}_i(t_j), \bm{\alpha}, \bm{\epsilon}}{\hat{O}_{\bm{\eta}}}{\bm{x}_i(t_j), \bm{\alpha}, \bm{\epsilon}}
\label{eq:expecatation}
\end{equation}
where $\bm{\eta}$ is a parameter vector $(\eta_0, \eta_1, \ldots, \eta_g)$. $\hat{O}_{\bm{\eta}}$ is permitted to be any $n$-qubit Hermitian observable and the precise choice can be considered an inductive bias of the model. In this work we choose $\hat{O}_{\bm{\eta}} = \eta_0 I - \frac 1n \sum_{i=1}^{n} \eta_i \hat{\sigma}_z^i$ where $\eta_i=1$ for $i\geq 1$ is fixed in the forthcoming parameter optimization but $\eta_0$ is variable within $-1 \leq \eta_0 \leq 1$. Next, we take the \textit{classical} expectation value of the square of Eq. \ref{eq:expecatation}, drawing $\bm{\epsilon} \sim \mathcal{N}(\bm{\mu}, \bm{\sigma})$ and re-scale by $L$
\begin{equation}
    C_1(\bm{x}_i(t_j), \bm{\theta}) := \frac{ \underset{\bm{\epsilon} \sim \mathcal{N}(\bm{\sigma}, \bm{\mu})}{\mathbb{E}} [\Omega^2 \left(\bm{x}_i(t_j), \bm{\alpha}, \bm{\epsilon}, \bm{\eta} \right)]}{L}
    \label{eq:C1}
\end{equation}
where $\bm{\theta} = [\bm{\alpha}, \bm{\mu}, \bm{\sigma}, \bm{\eta}]$ and $L$ is a factor chosen heuristically (in our specific implementation, $L=4$) in order to shrink the range of possible values to (0, 1). It should be mentioned that $L$ is not an essential part of the algorithm, and was only used to make the presentation more aesthetic. In practice, we estimate the right-hand side of Eq. \ref{eq:C1} by taking the mean value of $N_E$-many terms in the right-hand side brackets, where $N_E$ should be chosen either heuristically or once it becomes apparent that the differences in the mean value upon increasing $N_E$ are sufficiently small. With our explicit choice of $\hat{O}_{\eta}$, Eq. \ref{eq:C1} measures the average square distance between $\eta_0$ and $\braketmatrix{\bm{x}_i(t_j), \bm{\alpha}, \bm{\epsilon}}{\sum_{i=1}^{n}\hat{\sigma}_z^i}{\bm{x}_i(t_j), \bm{\alpha}, \bm{\epsilon}}$ for $N_E$-many sampled $\bm{\epsilon}$. Eq. \ref{eq:C1} is the fundamental building block of the forthcoming loss function and can be considered the cost contribution from the $i^{\text{th}}$ time series at the $j^{\text{th}}$ time point. Taking the average $\forall t \in B_{\mathcal{T}}$, we define the cost for the entirety of the $i^{\text{th}}$ time series as
\begin{equation}
    C_2(\bm{x}_i, \bm{\theta}) := \frac{1}{N_{\mathcal{T}}} \sum_{t_j \in B_{\mathcal{T}}} C_1(\bm{x}_i(t_j), \bm{\theta})
    \label{eq:C2} \enspace.
\end{equation}
Our task is to find model parameters that cluster all instances of the expression in Eq. \ref{eq:C2} about a given point, which, with our choice of $\hat{O}_{\bm{\eta}}$, is a clustering about a center at $\eta_0$. To achieve this, we minimize the cost function
\begin{equation}
    C(\bm{\theta}) :=  P_{\bm{\tau}}(\bm{\sigma}) + \frac{1}{2N_X}
    \sum_{\bm{x}_i \in B_X} C_2(\bm{x}_i(t), \bm{\theta})
    \label{eq:cost}
\end{equation}
such that 
\begin{equation}
\bm{\theta^{\star}} := \text{argmin}_{\bm{\theta}}[C(\bm{\theta})]
\label{eq:minimization}
\end{equation}
where the optimal parameters $\bm{\theta}^{\star}$ are found with a minimization routine on a classical computer where each new step of the minimizer estimates $C(\bm{\theta})$ with new mini-batches $B_X$ and $B_\mathcal{T}$, terminating at a fixed number of epochs (equivalently, mini-batch iterations) or when $C(\bm{\theta})$ falls below a certain tolerance. $P_{\bm{\tau}}(\bm{\sigma})$ is an explicit regularization function with $\gamma$-many hyperparameters $\bm{\tau} = \{\tau_1, \tau_2, \ldots, \tau_{\gamma}\}$ designed to penalize large entries of $\bm{\sigma}$. While the positive domain of any sigmoidal function can be used, throughout this work, we take
\begin{equation}
P_{\bm{\tau}}(\bm{\sigma}) := \frac{1}{\pi Q} \sum_{m=1}^{Q} \arctan(2 \pi \tau_m |\sigma_m|)  
\end{equation}
where $\tau_1=...=\tau_{Q-1}=\tau$ for a single contraction hyperparameter $\tau$. With this choice, $C(\bm{\theta})$ assumes values in the interval $(0, 1)$. The motivation for including a $P_{\bm{\tau}}(\bm{\sigma})$ regularization is that for a fixed $\bm{\alpha}$, letting $b$ be the length of $\bm{\epsilon}$, the mapping $f_{\bm{\alpha}}: R^b \rightarrow U(b)$ defined by $\bm{\epsilon} \mapsto W(\bm{\alpha})^{\dagger}e^{-iM(\bm{ \epsilon})}W(\bm{\alpha})$ gives rise to a set  $\mathrm{Im}(f_{\bm{\alpha}})$ of unitary matrices which we wish to controllably restrict. The distribution $\mathcal{N}(\bm{\sigma}, \bm{\mu})$ induces a probability distribution on $\mathrm{Im}(f_{\bm{\alpha}})$, which, as is indicated in the SM \cite{supplement}, by minimizing $\bm{\sigma}$ we contribute to increasing $C_2(\bm{y}, \bm{\theta}^{\star})$ for time series outside of the training set. \par

After a sufficiently good approximation of $\bm{\theta}^{\star}$ is obtained, it is possible to evaluate the anomaly score of an unseen time series $\bm{y}$ by letting
\begin{equation}
    a_X[\bm{y}] := |2C(\bm{\theta}^{\star}) -2P_{\bm{\tau}}(\bm{\sigma})  - C_2[\bm{y}, \bm{\theta}^{\star}]|.  
\label{eq:series_anomaly_score}
\end{equation}
Time series are classified as anomalous whenever the threshold $a_X[\bm{y}] > \zeta$ is exceeded. Otherwise, time series are classified as normal. If the data are available, in practice, $\zeta$ is chosen to maximize a performance metric (such as $F_1$ score or balanced accuracy score) using a validation set of labelled anomalous time series. \par

\section*{III. \quad Demonstration}

\subsection*{A \quad Synthetic univariate time series}

 \begin{figure}
    \centering
    \includegraphics[width=\linewidth]{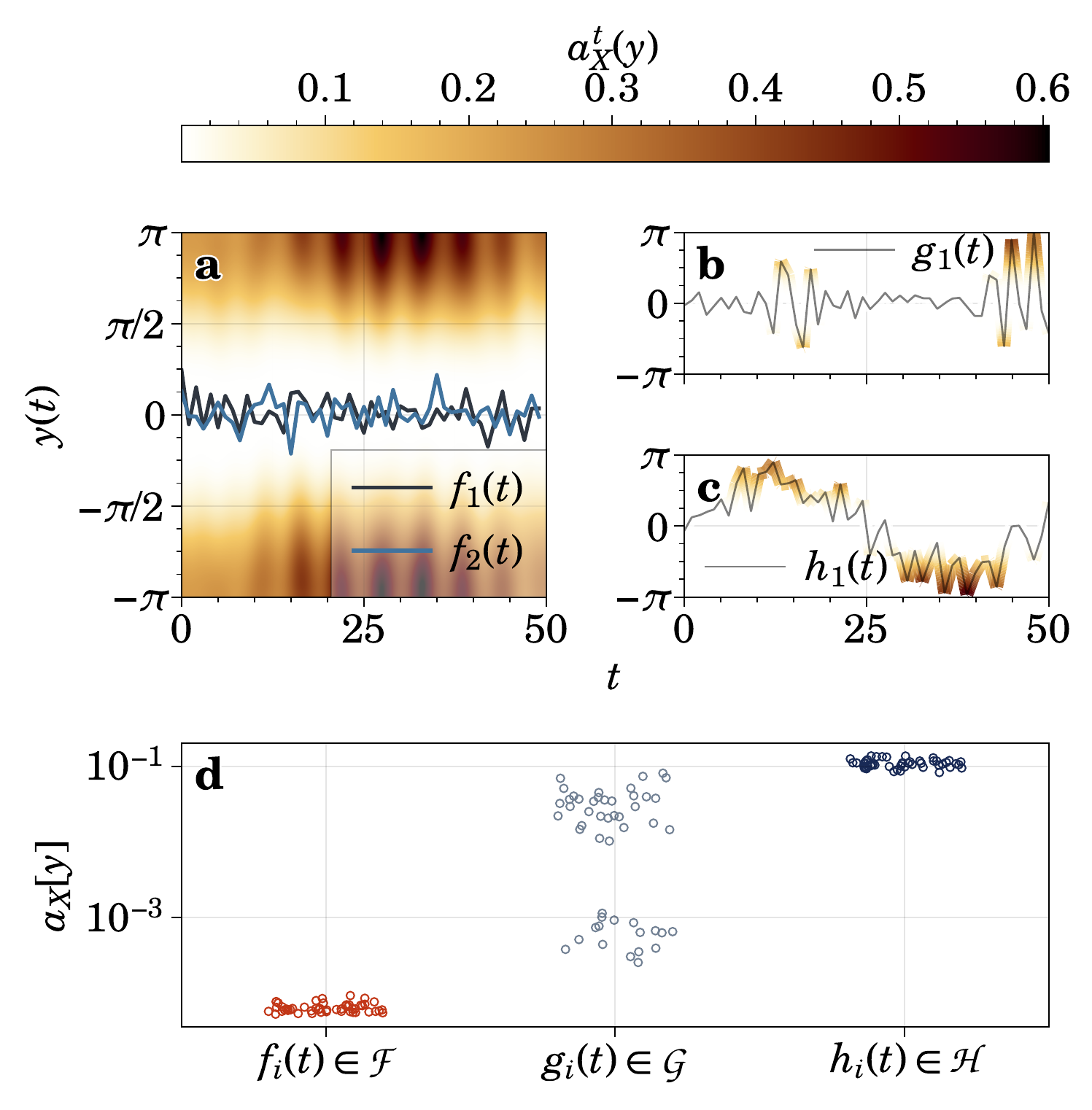}
    \caption{QVR applied to synthetically generated univariate training and testing data. (a) A heat-map of the time-resolved anomaly score $a^t_X(y)$. Two testing signals $f_1(t)$, $f_2(t) \in \mathcal{F}$ (generated using the same mechanism as the training data) are over-plotted and are seen in the region of low $a^t_X(y)$. (b-c) Example signals from the test sets $\mathcal{G}$ and $\mathcal{H}$ respectively, colored according to $a^t_X(y)$. $\mathcal{G}$ contains noisy signals with randomly inserted \textit{anomaly spikes} and $\mathcal{H}$ contains noisy $\sin(t)$ signals. (d) The anomaly scores $a_X[y]$ for each time series in the sets $\mathcal{F}$, $\mathcal{G}$ and $\mathcal{H}$. The y-axis has a logarithmic scale.}
    \label{fig:univariate_case}
\end{figure}

Here we shall demonstrate, using a didactic (and toy) example, how our classification algorithm provides results that are intuitively desirable. That is, the time series that are classified as normal are the ones we would intuitively consider as such. We shall use three sets of time series where the first set consists of elements that are similar to the original training data while the other two consist of time series that are ``intuitively different" than what's in the training set. Our goal will be to compute the anomaly score at each time point for each time series (this will be defined next) and to demonstrate that the time series that are intuitively similar to the training data will give rise to anomaly scores that are overall lower compared to the other time series. \par

The training data set $X$ contains 50 instances of noisy synthetic time series, each consisting of 50 time points where the values at each point are drawn from a normal distribution centered on the origin. The states $|x_i(t), \bm{\alpha}, \bm{\epsilon} \rangle$ are prepared with angle embedding on a 2-qubit circuit such that $U[(\bm{x}_i(t_j)] = R_y[x^1_i(t_j)] \otimes I$ (i.e, a $y$-rotation gate with an angle equal to $x^1_i(t_j)$ is applied to the first qubit) and $W(\bm{\alpha})$ is chosen to be a single layer of a circuit corresponding to the Ans{\"a}tz proposed in~\cite{Schuld2020}. All quantum expectation values are calculated exactly using the \texttt{lightning.qubit} quantum circuit simulator from the \texttt{Pennylane} QML package \cite{Bergolm2018}. We take $N_E=10$, $N_\mathcal{T} = 10$ and $N_X = 5$ and obtain $\bm{\theta}^{\star}$ from the run achieving the lowest $C(\bm{\theta})$ over 300 classical optimization runs each with 1000 mini-batch iterations, starting with randomly initialized model parameters. Further information regarding this process is given in the SM \cite{supplement}. \par

Using the trained model, we analyse three synthetic testing data sets: $\mathcal{F}$, $\mathcal{G}$ and $\mathcal{H}$. Each set contains the same number of instances and time points as $X$. Points $f_i(t) \in \mathcal{F}$ are generated using the same process as $X$ (although, $f_i(t) \notin X$). Points $g_i(t) \in \mathcal{G}$ are generated by starting with the same mechanism underlying $X$ and inserting \textit{anomalous spikes} of pseudo-random duration, frequency and amplitude. Lastly, $h_i(t) \in \mathcal{H}$ are generated by adding $\sin(t)$ to $f_i(t)$. Bold face is dropped on $f_i(t)$, $g_i(t)$ and $h_i(t)$ to denote that they are univariate time series. Figure \ref{fig:univariate_case}(a) provides insight into what is learnt in the training process. That is, upon examining the time-resolved anomaly score $a_X^t(\bm{y}):=|2C(\bm{\theta}^*) - 2P_{\bm{\tau}(\bm{\sigma})} - C_1(\bm{y}(t), \bm{\theta}^*)|$ on a fine grid, time series in $\mathcal{F}$ [see black and blue lines on Fig. \ref{fig:univariate_case}(a)] appear in areas where $a_X^t(y) \approx 0$ for all $t$ compared to series in $\mathcal H$ and $\mathcal G$. This accumulates to low $a_X[f_i]$ $\forall{i}$ as shown by the red markers on Fig. \ref{fig:univariate_case}(d). This is intuitive since $\mathcal{F}$ is generated with the same mechanism as $X$. The same cannot be said for series in $\mathcal{G}$ and $\mathcal{H}$. Figure \ref{fig:univariate_case}(b) and  \ref{fig:univariate_case}(c) each show an exemplar time series in $\mathcal{G}$ and $\mathcal{H}$, respectively, and are colored by the value of $a_X^t(y)$ (as indicated by the colorbar at the top of Fig. \ref{fig:univariate_case}). Common to both cases is the accumulation of time-resolved anomaly score at high amplitude regions of the signals. For series in $\mathcal{G}$, these regions are sporadic with randomized duration, which, as can be seen from the grey markers on Fig. \ref{fig:univariate_case}(d), leads to a spread in $a_X[y]$ over \textit{three orders of magnitude}. However, even the series achieving the lowest score still well exceeds the score of any series in $\mathcal{F}$. The sinusoidal component for series in $\mathcal{H}$ provide high amplitude regions of predictable duration, which, as is seen from the blue markers on Fig. \ref{fig:univariate_case} translates to high anomaly scores with a spread similar to $\mathcal{F}$. In a real setting, a threshold $\zeta$ would be set using a validation data set and classifications would be made. We note, however, that in our simple didactic case, there exist $\zeta$ values which perfectly separate $\mathcal{F}$ from $\mathcal{G}$ and $\mathcal{H}$, giving rise to perfect accuracy scores. This is unlikely to be achievable with real-world data sets. \par

\subsection*{B \quad Multivariate cryptocurrency time series}

Now, motivated by the multitude of applications for TAD sought after by central banks and other financial institutions, we tackle a more realistic problem with multivariate time series. That is, we use the QVR algorithm to study the behaviour of trading signals on the Binance crpytocurrency exchange \cite{binance2022} in a fixed time window after large amounts of Bitcoin (BTC) or United States Dollar Tether (USDT) are transferred from Binance exchange wallets to external wallets or vice versa. Such movements are noteworthy because they suggest a plausible intent to buy or sell a large number of coins on the exchange, which on a fixed-supply market like BTC, can strongly manipulate the market price or, at the very least, significantly affect trading dynamics on the exchange. Entities with holdings of cryptocurrency sufficient to drive these manipulation events are known as \textit{whales}. These whales, however, \textit{cannot} make these transactions privately. Inherent to many blockchain technologies (including BTC and USDT) is the recording of all transactions on a public ledger. This means whale transactions can be tracked and traders can be alerted of this using a service such as \texttt{Whale Alert} \cite{whale_alert_web2022}, now used to study trader behaviour impact on cryptocurrency markets~\cite{saggu2022intraday}. The problem we tackle is now more precisely defined: in some period of time around such an alert, is market activity detected as anomalous by QVR when a model is trained with time series following periods with no alerts? \par

% \VS{(double check data generation rules)}

% \VS{(rewrite for clarity)}

In our experiments, we extract time series windows from Binance with starting points coincident with different conditions reported by \texttt{Whale Alert}. The most broad condition is that an alert was given for either a large transaction of BTC or USDT on the blockchain between Binance and non-Binance wallets, resulting in two large data sets $\mathcal{B}$ and $\mathcal{U}$, respectively. Each contains 1900 time series instances. These two data sets are pruned (see SM for details \cite{supplement}) and further subdivided into six subsets: $\mathcal{U}_{\pm}$, $\mathcal{B}_{\pm}$, $\tilde{\mathcal{U}}_{+}$, $\tilde{\mathcal{U}}_{-}$, $\tilde{\mathcal{B}}_{+}$, $\tilde{\mathcal{B}}_{-}$. Here, the subscripts $+$ or $-$ indicate that transactions were into or out of Binance exchange wallets, respectively, while the subscript $\pm$ means transactions in both directions are permitted. Additionally, the tilde cap $\sim$ designates data sets where more than one large transaction of BTC or USDT were made during the time window under consideration. Those without the tilde cap have only a single large transaction. Next, training data set $X$ and normal data set $\mathcal{N}$, $X \cap \mathcal{N} = \varnothing$, contain time-series where no alerts were made in the time window. This is considered normal behaviour. Lastly, $\mathcal{V}$ is a validation data set containing equal numbers of series arising from the same conditions as sets $\tilde{\mathcal{U}}_{+}$, $\tilde{\mathcal{U}}_{-}$, $\tilde{\mathcal{B}}_{+}$, $\tilde{\mathcal{B}}_{-}$, but with no common elements. For each of these data sets, we consider up to three time series defined over the window: the mean deviation of the BTC-USDT price, the cumulative trading volume and the cumulative taker buy base asset (TBBA) volume. Models trained using only the first two series are called bivariate models and those trained using all three series are called trivariate models. Further information on the cryptocurrency data sets including their construction and pre-processing can be found in the SM \cite{supplement}.

\begin{figure}
    \centering
    \includegraphics[width=\linewidth]{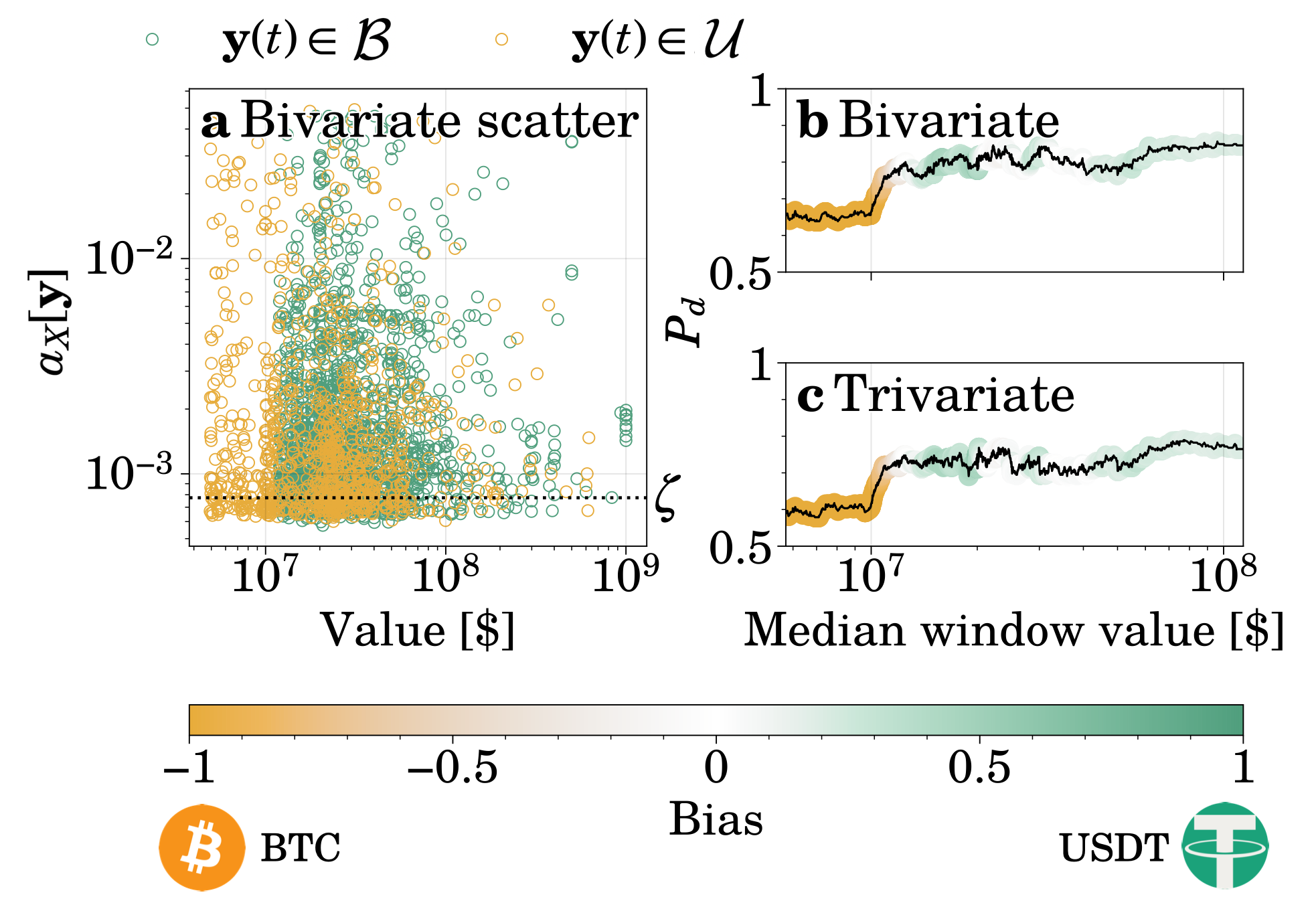}
    \caption{Quantifying anomalous cryptocurrency time series behaviour on the large $\mathcal{U}$ and $\mathcal{B}$ data sets using QVR. (a) A scatter plot of anomaly scores from the bivariate model versus the USD value of BTC and USDT transactions reported by \texttt{Whale Alert}. The threshold $\zeta$ is marked. (b) and (c) The detection probability $P_d = P(a_X[\bm{y}] > \zeta)$ within a moving window of size $W=285$ and step $s=1$ for the bivariate and trivariate models, respectively. Lines are colored according to the lower color bar and represent the proportion of transactions in the window that were on the BTC or USDT blockchains.}
    \label{fig:btc_vs_usdt}
\end{figure}

\begin{figure*}
    \centering
    \includegraphics[width=\linewidth]{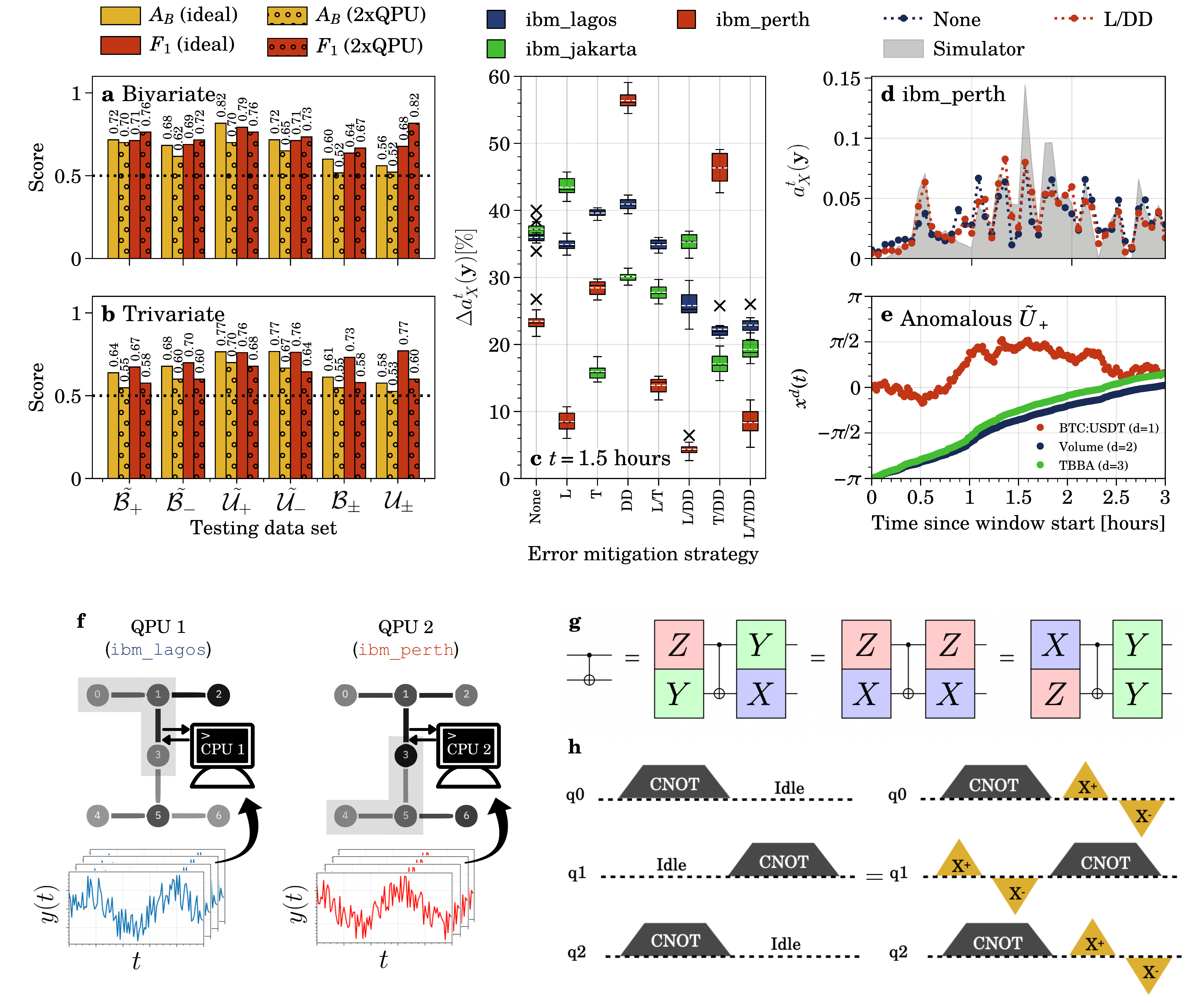}
    \caption{Experiments on 7-qubit superconducting transmon QPUs. (a) and (b) $F_1$ and balanced accuracy scores ($A_B$) for the bivariate and trivariate series, respectively. Scores are given for an ideal quantum simulator and for a dual QPU strategy as shown in (f). (c) The anomaly score error $\Delta a_X^t(y)$ compared with an ideal QPU for various error mitigation strategies: L = noise-aware layout, T = Pauli twirling, DD = dynamical decoupling. Forward slashes indicate their use in combination. (d) The time-resolved anomaly score $a_X^t(\bm{y})$ for a selected $\bm{y}(t) \in \tilde{\mathcal{U}}_+$ on \texttt{ibm\_perth}, with and without L/DD error mitigation. (2) The trivariate time series corresponding with (d). (f) The dual QPU strategy used in (a) and (b). Half of the time series are sent to one QPU and half to another, sped up by low-latency CPU interaction. Grey highlighted areas over nodes of the QPU graphs are exemplar noise-aware layouts of three qubits. (g) Exemplar gate identities used in Pauli twirling. (h) Illustration of an \texttt{X2pm} DD pulse sequence.}
    \label{fig:2col}
\end{figure*}

Similar to the univariate synthetic case, classical data is encoded with angle embedding such that $U[\bm{x}_i(t_j)] = R_y[x_i^1(t_j)] \otimes R_y[x_i^2(t_j)] \otimes \ldots R_y[x_i^d(t_j)]$. We choose $n=d$ such that the bivariate and trivariate models are embedded in two- and three-qubit systems, respectively. A three layer circuit corresponding to the same Ans\"{a}tz as before \cite{Schuld2020} is used to represent $W(\bm{\alpha})$. Also like in the univariate synthetic case, quantum expectation values needed in training are calculated exactly using the \texttt{lightning.qubit} quantum circuit simulator from the  \texttt{Pennylane} QML package \cite{Bergolm2018}. Taking $N_X = N_E = N_{\mathcal{T}} = 10$, both models are now trained 50 times for 2000 mini-batch iterations at which point the models are used evaluate $a_X[\bm{y}]$ for the validation set $\mathcal{V}$ and normal set $\mathcal{N}$ to tune the threshold $\zeta$ to maximize the balanced accuracy score $A_B$. This procedure is repeated for all 50 training instances to find optimal models; those which produce the largest $A_B$. For the optimal bi/trivariate models we obtain $A_B=0.82/0.77$ respectively. The optimal models are now used to evaluate $a_X[\bm{y}]$ for all of the testing data sets, and, with the optimal $\zeta$, evaluate $A_B$ and $F_1$ scores using $\mathcal{N}$ and the previously defined subsets of $\mathcal{U}$ and $\mathcal{B}$. \par

The results are shown in Fig. \ref{fig:btc_vs_usdt} and on the solid bars of Fig. \ref{fig:2col}(a) and \ref{fig:2col}(b). In Fig. \ref{fig:btc_vs_usdt}(a), we observe that a large number of time series in $\mathcal{U}$ and $\mathcal{B}$ are identified as anomalous by the bivariate model, but there is no clear relationship between the value of the transaction and the anomaly score. We do, however, see that the probability of a time series being classified as anomalous in a value window of fixed size, $P_d = P(a_X[\bm{y}] > \zeta)$, is a non-monotonic but broadly increasing function of the median window value [Fig. \ref{fig:btc_vs_usdt}(b) and \ref{fig:btc_vs_usdt}(c)]. This is true for the bivariate and trivariate models but detection probability is generally higher for the former. This trend is also complicated by the fact that lower value transactions tend to be dominated by BTC [see orange regions near origin of Fig. \ref{fig:btc_vs_usdt}(b) and \ref{fig:btc_vs_usdt}(c)] while larger transactions are either biased towards USDT or evenly split between the two [the green and green/white regions on Fig. \ref{fig:btc_vs_usdt}(b) and \ref{fig:btc_vs_usdt}(c)]. We therefore conclude that $P_d$ and transaction value in USDs are positively correlated and both models are more likely to identify market behaviour following large USDT transactions as anomalous. \par

Examining Fig. \ref{fig:2col}(a) and \ref{fig:2col}(b), we first note that $\mathcal{B}_{\pm}$ and $\mathcal{U}_{\pm}$ are only classified successfully marginally more often than a random guess. This can be attributed to the relatively low transactions of BTC and USDT within these data sets. Conversely, $A_B$ is close to $0.8$ for $\tilde{\mathcal{U}}_{+}$ for both models. As is seen in Fig. \ref{fig:btc_vs_usdt}, transactions in these data sets can be $\sim \$1$ billion suggesting that large influxes of USDT into exchange wallets produces market behaviour strongly deviant from normal conditions. Moving from the bivariate to trivariate model, there are small differences in performance, mainly manifesting in better $A_B$ and $F_1$ for $\tilde{\mathcal{U}}_-$ and $\mathcal{B}_{\pm}$ but slightly degraded performance for $\tilde{\mathcal{U}}_+$ and $\tilde{\mathcal{B}}_+$. In summary, transactions of BTC or USDT to/from exchange wallets leads to market behaviour which, according to QVR, are frequently detected as anomalous. We find that the most likely time series to produce abnormal market behaviour are those windows after a large transaction of USDT from an external wallet to a Binance exchange wallet.

\subsection*{C \quad Experiments on superconducting transmon quantum processing units}

We now study how well these findings translate when QVR is deployed to real hardware. Namely, we perform validation (including the re-optimization of $\zeta$) and testing once more with the bivariate and trivariate models using IBM's \texttt{Falcon r5.11H} family of superconducting transmon QPUs \cite{ibmq2022} (key QPU specifications are given in the SM \cite{supplement}). Given the processor speeds of present QPUs of this type ($\sim 1000$ circuit layer operations per second; CLOPS \cite{Wack2021}), classical-quantum communication latency (primarily network latency for accessing cloud QPUs), provider-set caps on the quantum resources usable in one execution on the cloud and the large number of expectation values that must be estimated ($\approx 65,000$ in total), steps must be taken to make these computations possible in a reasonable time. Our first step is to adopt a dual QPU strategy whereby half of the time series in each of the data sets are sent to \texttt{ibm\_lagos} and the other half to \texttt{ibm\_perth} (see Fig. \ref{fig:2col}(f), also showing the qubit connectivity graph of both QPUs). This parallelization leads to a trivial 2x speedup. We note that this is not the optimal approach as a maximum of four qubits are idle on these seven qubit devices. Using distributed computing strategies like the Quantum Message Passing Interface (QMPI) \cite{Haner2021}, these idle qubits can be utilized to achieve greater speedups. We do not adopt this approach since we later experiment with noise-aware qubit layouts [see grey shaded areas on Fig. \ref{fig:2col}(f)] which benefit from being able to choose from all of the relevant sub-graphs on the chip. The second step is to change the way in which data containing circuit descriptions are sent to the QPUs. That is, the testing step of QVR is deployed to CPUs ``near" the QPUs using \texttt{Qiskit Runtime} which (i) strongly reduces classical-quantum communication latency and (ii) allows for more circuits to be executed in the same period of time since quantum resources are now capped by total runtime rather than a number of shots. These steps are outlined in Fig. \ref{fig:2col}(f) and its caption. We also work at a smaller number of shots, using 256 and 512 shots per expectation value estimation for the bivariate and trivariate models, respectively. \par 

The results are displayed as the circle-filled bars on Fig. \ref{fig:2col}(a) and \ref{fig:2col}b. Compared with the trivariate model, the performance of the bivariate model is fairly faithful to the ideal (noise-free) result in the presence of real noise. Interestingly, in the bivariate case, $F_1$ scores for the $\mathcal{U}_{\pm}$ and $\mathcal{B}_{\pm}$ data sets actually improve on QPU. When measured against the almost random classification accuracy however, this increased $F_1$ is an artefact of the re-optimized $\zeta$ returning a near-perfect recall score (correctly classifying $\bm{y} \in \mathcal{N}$) at the cost of low precision (incorrectly classifying $\bm{y} \in \mathcal{U}_{\pm}$ and $\bm{y} \in \mathcal{B}_{\pm}$). We observe that the general effect of noise is to compress the range of achievable $a_X[\bm{y}]$. This is the expected effect since decoherence will eventually lead to $\langle \hat{\sigma}^i_z \rangle \rightarrow 0.5$ regardless of the time series in question. This is a manifestation of the well-known noise-induced barren plateaus phenomenon \cite{Wang2021}, where our plateaus appear in the anomaly score landscape. This landscape flattening coupled with shot noise means that \textit{close call} time series (i.e, those with anomaly scores close to $\zeta$ for an ideal QPU) are classified incorrectly as an increasing function of both noise sources. \par

Faced with the larger errors present for the trivariate model, we conduct an investigation into the use of error mitigation techniques; widely considered a necessary ingredient for realizing ``useful" quantum advantage on NISQ hardware \cite{Song2019, Youngseok2021}. To begin, we probe the error in the time-resolved anomaly score $\Delta a^t_X(\bm{y})$ at a single point in time ($t = 1.5$ hours) using an arbitrarily selected time series $\bm{y} \in \tilde{\mathcal{U}}_+$. That is, using three different error mitigation strategies and their combinations: Noise-aware layouts \cite{Li2021, Mapomatic2022}, Pauli twirling (also known as Pauli conjugation \cite{Cai2019, Cai2020}) and dynamical decoupling (DD) \cite{Viola1999, Ezzell2022} we compute $\Delta a^t_X(y)$ for ten repeats on \texttt{ibm\_lagos}, \texttt{ibm\_perth} and \texttt{ibm\_jakarta}, increasing the number of shots to 4096 to strongly reduce the level of shot noise (necessary to resolve would-be performance improvements from error mitigation over the level of the stochastic error from shot-noise). A noise-aware layout is the lowest noise sub-graph on the wider machine connectivity graph for a given compiled quantum circuit. We perform this step using \texttt{mapomatic} (\texttt{v0.7}) \cite{Mapomatic2022}. Pauli twirling aims to average out coherent error by implementing simple gate identities for high-error 2-qubit gates. This is shown in Fig. \ref{fig:2col}(g). To achieve the twirled result, we take the average expectation value of 10 randomly twirled circuits. Dynamical decoupling uses controlled pulse sequences to approximately cancel system-environment coupling and the resulting decoherence. We use an \texttt{X2pm} sequence which is illustrated in Fig. \ref{fig:2col}(h). For a summary of these techniques using IBM QPUs, we refer to reader to Ref. \cite{Youngseok2021}. \par

As can be seen in Fig. \ref{fig:2col}(c), using a combination of all of these techniques [L/T/DD on Fig. \ref{fig:2col}(c)] reliably reduces anomaly score errors for all three QPUs, but is not observed as the single best strategy for any given chip. We see that it is also possible for some error mitigation techniques to give rise to an increase in error. This further highlights the importance of performing these experiments as the optimal strategy is not trivial to determine, depending sensitively on the precise error characteristics of the device. Looking at the results for \texttt{ibm\_perth} [red box-plots on Fig. \ref{fig:2col}(c)], we see that $\Delta a^t_X(y)$ can be strongly reduced to $\approx 4\%$ by leveraging a combination of a noise-aware layout and DD. Encouraged by this low error, we determine whether it persists at other $t$-points as shown in Fig. \ref{fig:2col}(d) above the trivariate time series in question [Fig. \ref{fig:2col}(e)]. It is immediately clear that the error [see the difference between the colored lines and grey shaded outline on Fig. \ref{fig:2col}(e)] is strongly time dependent. Despite this, the sum of the absolute difference between the QPU result and ideal simulator is reduced by $\approx 14$\% when the error mitigation strategy is used. This is valuable early evidence that the performance of QML can be robustly improved on real hardware should one perform careful testing on the choice of error mitigation strategy. \par

\section*{IV. \quad Conclusions}

In summary, we have proposed and demonstrated the execution of a low circuit depth QML algorithm for TAD: \textit{Quantum Variational Rewinding} or, QVR. QVR learns a distribution of parameterized unitary time-devolution operators able to \textit{rewind} time series representing the normal behaviour of a system/process. Anomalous series are those which cannot be rewound by the learnt normal model. After providing a simple and didactic use case on synthetic univariate time series, we demonstrated that QVR can be used to detect anomalies in realistic multivariate time-series. That is, in an investigation into trader behaviour on cryptocurrency markets in the presence of \texttt{Whale Alerts}, QVR identified that market behaviour was detectably different when large quantities of USDT or USDT flowed to/from exchange wallets from/to external wallets, as identified by a \texttt{Whale Alert}. Anomalous behaviour was identified most frequently when (i) transactions were the largest and (ii) where transactions of USDT were into exchange wallets. Lastly, using a trifecta of parallel QPU execution, near-QPU classical resources and advanced error mitigation techniques, we showed the that QVR is resistant to noise when executed on present NISQ hardware. These techniques, in tandem, will allow for ever-more useful QML workflows to become realizable using NISQ QPUs. While we have suggested possible areas where QVR could provide some advantages over classical techniques \cite{Caro2022, Aharonov2022, Huang2021}, adopting the stance taken in Ref. \cite{Schuld2022}, we suggest that some case-by-case heuristic advantages over classical methods \textit{could} appear as the scale and quality of QPUs improves. Given this, at this stage in the maturity of NISQ hardware, focusing solely on the topic of advantage may not necessarily be the best approach \cite{Schuld2022}. Demonstrations in this work can then be considered proof-of-concept examples in fundamental quantum machine learning research.

\section*{Acknowledgements}

We acknowledge the use of IBM Quantum services for this work. The views expressed are those of the authors, and do not reflect the official policy or position of IBM or the IBM Quantum team. \par

The views expressed in this paper are solely those of the authors and may differ from official Bank of Canada views. No responsibility for them should be attributed to the Bank. \par

We acknowledge the use of the \href{https://github.com/qiskit-research/qiskit-research}{Qiskit-Research} repository in this work. \medskip

\section*{Data and Code availability}

All of the pre-processed time series data sets used in this work and an explicit code implementation of QVR can be found at either

\begin{enumerate}
    \item The public \href{https://github.com/AgnostiqHQ/QuantumVariationalRewinding}{GitHub repository}
    
    \item The \href{https://doi.org/10.5281/zenodo.7258627}{Zenodo mirror}
\end{enumerate}

% [I've commented this out because it bring up too many questions for referees. i.e, why would this be better for multidimensional series? Why would you use this instead of classical?]interested in anomaly detection for many applications. One such application is the high volumes of data received from financial institutions. Currently, this data is analyzed by a classical machine learning detection algorithm. Other methods, such as those mentioned in this paper, are useful as alternatives to the current methodologies, especially relating to high-dimensional, multivariate time-series. \par

% \textcolor{cyan}{Jack: For the rest of the demonstration section we need to show the algorithm being trained on real QPUs using the data given by BOC. This will create one two-column figure with $C(\bm{\theta})$ varying with epochs with and without using error mitigation on \texttt{QiskitRuntime}. The figure will also include a classification accuracy plot (probably a histogram).}

% \textcolor{cyan}{Jack: Write conclusion highlighting that (i) the algorithm currently works on real hardware, but will get better with improving hardware (ii) the finance application was a relevant and real-world problem being solved with QML.}

% \nocite{*}
% \bibliographystyle{apsrev4-1

% Produces the bibliography via BibTeX.

\bibliography{ms}

\clearpage
\newpage
\mbox{~}
\onecolumngrid
\setcounter{equation}{0}
\setcounter{figure}{0}
\setcounter{table}{0}
\setcounter{page}{1}
\setcounter{secnumdepth}{2}

\makeatletter
\renewcommand{\theequation}{S\arabic{equation}}
\renewcommand{\thefigure}{S\arabic{figure}}
\renewcommand{\bibnumfmt}[1]{[S#1]}
\renewcommand{\citenumfont}[1]{S#1}
\vspace*{\fill}
\begin{center}
\textbf{\large Supplemental Material: Quantum Variational Rewinding for Time Series Anomaly Detection}
\end{center}
\section*{Overview}

This document contains all of the necessary supplemental material for the article ``Quantum Variational Rewinding for Time Series Anomaly Detection". We have structured this document as follows: Section \ref{sec:QVRindetail} provides a closer look at Quantum Variational Rewinding (QVR), providing algorithm pseudocode (Section \ref{subsec:algpseudo}), details on where the reader can find explicit Python code (Section \ref{subsec:codeavail}), a tabulation of all of the model parameters/hyperparameters with descriptions for ease of reference (Section \ref{subsec:components}) and neccessary conditions for the success of QVR (Section \ref{subec:formalism}). Section \ref{sec:univariate} provides further numerical experiments on synthetic univariate time series. Using the didactic case from the main article, we investigate the role of different model parameters (Section \ref{subsec:univariate_expec} and \ref{subsec:univariate_mu_sigma}) and study the effectiveness of different derivative-free optimizers used in training the model (Section \ref{subsec:univariate_derivative_free}). We also include another study where noisy univariate sinusoids are learnt (Section \ref{fig:noisy_sinusoids}). This study shows that QVR produces an anomaly scoring function which is not related to a simple euclidean distance between training time series and testing time series. Section \ref{sec:multivariate_synthetic_static} discusses the role of the penalty term $P_{\bm{\tau}}(\mathbf{\sigma})$. First, the relationship of $P_{\bm{\tau}}(\mathbf{\sigma})$ and the concept of \textit{latent spaces} is discussed (Section \ref{subsec:latent}). Then with minimal modification, we show that the algorithm is also suitable for static anomaly detection (i.e, data sets without temporal dependence; Section \ref{subsec:static}). Using this variation of the algorithm, we present a study of the effects of varying the penalty function hyperparameter $\tau$. Section \ref{sec:realMultivariate} describes in more detail the cryptocurrency time series use case from the main article. This includes details about the data set, its generation and its availability (Section \ref{subsec:cryptodataset}) as well presenting the training performance for the bivariate and trivariate models (Section \ref{subsec:training_crypto}). Lastly, further details about the superconducting transmon quantum processing unit (QPU) runs (\ref{subsec:transmon}) are given including specifications of each machine used and an overview of our custom \texttt{Qiskit Runtime} program. 
\vspace*{\fill}

\newpage

\section{QVR in detail \label{sec:QVRindetail}}

\subsection{Algorithm pseudocode \label{subsec:algpseudo}}
We summarize the variation of QVR used in the main article below using algorithm pseudocode. There are two steps detailed: training (for a single epoch) and testing (for a single unseen time series $\bm{y}$). In the below, EXP denotes $\{1, 2, \ldots, N_E \}$. Please see the main article for variables not defined below.

\begin{algorithm}[H]
\caption{Training Phase}
\label{alg:training}

\textbf{Input:} A set of $d$-dimensional time series $X=\{\bm{x}_i : i=1,...,m\} \subseteq R^{p\times d}$ (each with $p$-many time points $\{t_1,...,t_p \}$), optimization routine, EPOCH, EXP, a penalty function $P$, two Ans\"{a}tze $U$ and $W$ and contraction hyperparameters $\boldsymbol{\tau}=(\tau_1,..., \tau_Q)$.

\begin{algorithmic}
\medskip
\State Set initial values for the parameters $\boldsymbol{\theta}=[\boldsymbol{\alpha}, \boldsymbol{\mu}, \boldsymbol{\sigma}, \boldsymbol{\eta}]$.

\For{$j=1$ to EPOCH}

\State Choose random batches $B_X \subseteq X$ and $B_T \subseteq \{t_1,...,t_p\}$ and let $N_X=|B_X|$ and $N_T=|B_T|$.

\For{$i=1$ to $B_X$}

\For{$k$ in EXP}

\State Choose a diagonal matrix $M(\boldsymbol{\epsilon})$ with eigenvalues $\epsilon_s$ chosen uniformly randomly with $\epsilon_s \sim \mathcal(\mu_s, \sigma_s)$

\For{$t_j$ in $B_T$}

\State Prepare the state embedding circuit $U[\bm{x}_i(t_j)]:=U[x_i^1(t_j)] \otimes... \otimes U[x_i^d(t_j)]$

\State Prepare the state $|\bm{x}_i(t_j) \rangle:=U[\bm{x}_i(t_j)]|0\rangle^{\otimes d}$

\State Prepare the circuit $W^{\dagger}(\bm{\alpha})D(\boldsymbol{\epsilon}, t_j)W(\bm{\alpha})=W^{\dagger}(\bm{\alpha})e^{-iM(\boldsymbol{\epsilon})t_j}W(\bm{\alpha})$

\State Prepare the state $|\bm{x}_i(t_j), \bm{\theta} \rangle:=W^{\dagger}(\bm{\alpha})D(\boldsymbol{\epsilon}, t_j) W(\bm{\alpha})|\bm{x}_i(t_j)\rangle$

\State Compute the single point cost $\Omega (\bm{x}_i(t_j), \boldsymbol{\alpha}, \boldsymbol{\epsilon}, \boldsymbol{\eta}):=\langle \bm{x}_i(t_j), \boldsymbol{\alpha}, \boldsymbol{\epsilon} | \eta_0 I- \frac{1}{n} \sum_{i=1} \eta_i \sigma_z^i| \bm{x}_i(t_j), \boldsymbol{\alpha}, \boldsymbol{\epsilon} \rangle$ 

\EndFor

\State Compute an instance of the single time series cost function $C_2(\bm{x}_i, \boldsymbol{\theta})^-:=\frac{1}{N_T} \sum_{t_j} \Omega^2 (\bm{x}_i(t_j), \boldsymbol{\alpha}, \boldsymbol{\epsilon},  \boldsymbol{\eta})$
\State using $\boldsymbol{\theta}=[\boldsymbol{\alpha}, \boldsymbol{\mu}, \boldsymbol{\sigma}, \boldsymbol{\eta}]$

\EndFor

\State Compute the intermediate cost function $C_2(\bm{x}_i, \boldsymbol{\theta})=E_{\epsilon_s \sim \mathcal{N}(\sigma_s, \mu_s)}[C_2(\bm{x}_i, \boldsymbol{\theta})^-]$

\EndFor

\State Compute the final cost function $C(\boldsymbol{\theta}):=\frac{1}{2N_X} \sum_{\bm{x}_i \in B_X} C_2(\bm{x}_i, \boldsymbol{\theta})+\frac{1}{\pi Q} \sum_m \arctan(2 \pi \tau_m |\sigma_m|)$

\State Run next step of the classical optimization routine

\State Update the parameters $\boldsymbol{\theta}=[\boldsymbol{\alpha}, \boldsymbol{\mu}, \boldsymbol{\sigma}, \boldsymbol{\eta}]$

\EndFor

\end{algorithmic}
\textbf{Output:} The optimized parameters $\boldsymbol{\theta}^*=[\boldsymbol{\alpha}^*, \boldsymbol{\mu}^*, \boldsymbol{\sigma}^*, \boldsymbol{\eta}^*]$ and the cost $C(\boldsymbol{\theta}^*)$.
\end{algorithm}

\begin{algorithm}[H]
\caption{Classification Phase}
\label{alg:classification}
\textbf{Input:} A $d$-dimensional time series $\bm{y} \in R^{p\times d}$, the optimized parameters ${\boldsymbol{\theta}}^*=[\boldsymbol{\alpha}^*, \boldsymbol{\mu}^*, \boldsymbol{\sigma}^*, \boldsymbol{\eta}^*]$, EXP, the Ans\"{a}tze $U$ and $W$ chosen in the training phase and the penalty function $P$ chosen in the training phase. \medskip
\begin{algorithmic}

\For{$k$ in EXP}

\State Choose a diagonal $M(\epsilon)$ with eigenvalues $\epsilon_s$ chosen randomly with $\epsilon_s \sim \mathcal(\mu_s^*, \sigma_s^*)$ 

\For{$t_j$ in $B_T$}

\State Prepare the state embedding circuit $U[\bm{y}(t_j)]:=U[y^1(t_j)] \otimes... \otimes U[y^d(t_j)]$ 

\State Prepare the state $|\bm{y}(t_j) \rangle:=U[\bm{y}(t_j)]|0\rangle^{\otimes d}$

\State Prepare the circuit $W^{\dagger}(\bm{\alpha}^*)D(\boldsymbol{\epsilon}, t_j) W(\bm{\alpha}^*)=W^{\dagger}(\bm{\alpha}^*)e^{-iM(\boldsymbol{\epsilon})t_j} W(\bm{\alpha}^*)$

\State Prepare the state $|\bm{y}(t_j), \bm{\theta}^* \rangle:=W^{\dagger}(\bm{\alpha}^*)D(\boldsymbol{\epsilon}, t_j) W(\bm{\alpha}^*)|\bm{y}(t_j)\rangle$

\State Compute the single point cost $\Omega (\bm{y}(t_j), \boldsymbol{\alpha}^*, \boldsymbol{\epsilon}^*, \boldsymbol{\eta}^*):=\langle \bm{y}(t_j), \boldsymbol{\alpha}^*, \boldsymbol{\epsilon}^* | \eta_0^* I- \frac{1}{n} \sum_{i=1} \eta_i^* \sigma_z^i| \bm{y}(t_j), \boldsymbol{\alpha}^*, \boldsymbol{\epsilon}^* \rangle$

\EndFor

\State Compute an instance of the single time series cost function $C_2(\bm{y}, \boldsymbol{\theta}^*)^-:=\frac{1}{N_T} \sum_{t_j} \Omega^2 (\bm{y}(t_j), \boldsymbol{\alpha}^*, \boldsymbol{\epsilon}^*,  \boldsymbol{\eta}^*)+P_{\bm{\tau}}(\boldsymbol{\sigma})$

\EndFor

\State Compute the single cost function $C_2(\bm{y}, \boldsymbol{\theta}^*)=E_{\epsilon_s \sim \mathcal{N}(\sigma_s^*, \mu_s^*)}[C_2(\bm{y}, \boldsymbol{\theta}^*)^-]$

\end{algorithmic}

\textbf{Output:} $a_X[\bm{y}] := |2C(\boldsymbol{\theta}^*)-2P_{\bm{\tau}}(\bm{\sigma})-C_2(\bm{y}, \boldsymbol{\theta}^*)|$ where $P_{\bm{\tau}}(\bm{\sigma})=\frac{1}{\pi Q} \sum_m \arctan(2 \pi \tau_m |\sigma_m|)$.

\end{algorithm}

\subsection{Code availability \label{subsec:codeavail}}

A Python implementation of QVR can be found in the public Github repository \url{https://github.com/AgnostiqHQ/QuantumVariationalRewinding} or in the file hosted by Zenodo (\url{https://doi.org/10.5281/zenodo.7258627}). The implementation breaks the algorithm down into modular parts and visualizes how these parts interact from a graph theory perspective. Then, the code goes through training, validation and testing using the bivariate model using the cryptocurrency time series case from the main article.

\subsection{Summary of algorithm components \label{subsec:components}}

Table \ref{tab:QVRComponents} contains the customizable components of QVR along with descriptions.

\begin{table}[h]
\centering
\begin{tabular}{@{}ll@{}}
\toprule
Component                    & Description and relationship with other elements of QVR                                                                                                                                                                                                                                                                                                                                                                                                                                                  \\ \midrule
$W(\bm{\alpha})$             & \begin{tabular}[c]{@{}l@{}}Parameterized unitary used in the eigendecomposition of $e^{-iH(\bm{\alpha}, \bm{\epsilon})}$ where $\bm{\alpha}$ and $\bm{\epsilon}$ are vectors\\ of free parameters. $\bm{\epsilon}$ is sampled from $\mathcal{N}(\bm{\mu}, \bm{\sigma})$ where $\bm{\mu}$ and $\bm{\sigma}$ \\ are also free parameters of the model. This can be chosen to be any parameterized quantum circuit.\end{tabular}                                                                            \\
                             &                                                                                                                                                                                                                                                                                                                                                                                                                                                                                                          \\
$D(\bm{\epsilon}, t_j)$      & \begin{tabular}[c]{@{}l@{}}Parameterized diagonal unitary used in the eigendecomposition of $e^{-iH(\bm{\alpha}, \bm{\epsilon})}$  where $\bm{\alpha}$\\ and $\bm{\epsilon}$ are vectors of free parameters.  $\bm{\epsilon}$ is sampled from $\mathcal{N}(\bm{\mu}, \bm{\sigma})$\\ where $\bm{\mu}$ and $\bm{\sigma}$ are also free parameters of the model. $t_j$ is a a time point. This can be chosen to be any \\ parameterized quantum circuit implementation of a diagonal unitary.\end{tabular} \\
                             &                                                                                                                                                                                                                                                                                                                                                                                                                                                                                                          \\
$U[\bm{x}_i(t_j)]$           & \begin{tabular}[c]{@{}l@{}}Unitary used to embed a time series $\bm{x}_i$ at a point $t_j$ into a quantum state $|\bm{x_i}(t_j) \rangle$. This is often referred to as \\ a ``quantum feature map". There are a large number of choices for such a unitary \cite{Schuld2021}.\end{tabular}                                                                                                                                                                                                                                 \\
                             &                                                                                                                                                                                                                                                                                                                                                                                                                                                                                                          \\
$P_{\bm{\tau}}(\bm{\sigma})$ & \begin{tabular}[c]{@{}l@{}}A penalty function designed to penalize large entries of the parameter vector $\bm{\sigma}$.  $\bm{\tau}$ is a vector of hyperparameters. \\ Any sigmoidal function can be used for the penalty.\end{tabular}                                                                                                                                                                                                                                                                 \\
                             &                                                                                                                                                                                                                                                                                                                                                                                                                                                                                                          \\
$\hat{O}_{\bm{\eta}}$        & A general $n$-qubit observable parameterized by the parameter vector $\bm{\eta}$.                                                                                                                                                                                                                                                                                                                                                                                                                        \\
                             &                                                                                                                                                                                                                                                                                                                                                                                                                                                                                                          \\
Classical optimizer          & \begin{tabular}[c]{@{}l@{}}The routine on a classical computer used to find $\bm{\theta^{\star}} := \text{argmin}_{\bm{\theta}}[C(\bm{\theta})]$. Performance will\\ also be affected by (1) the choice initial parameters $\bm{\theta}_{\text{init}}$,  (2) the time series and time point mini-batch sizes\\  $N_{X}$ and $N_{\tau}$, respectively and (3) the termination criteria of the optimization.\end{tabular}                                                                                  \\ \bottomrule
\end{tabular}
\caption{The customizable components of QVR.}
\label{tab:QVRComponents}
\end{table}

\subsection{On the success of QVR \label{subec:formalism}}

Below we isolate a property that implies the possible success of our algorithm and we suggest some basic and natural related problems.
\\
\\
\textbf{Definition}: Let $\epsilon>0$ and let $\Gamma=\{|\bm{x}_{i,j} \rangle\ : 1\leq i \leq m, 1\leq j \leq n \}$ be a set of states, all of which are of the same number of qubits. We say that $\Gamma$ is $\epsilon$-centered if there exists a Hamiltonian $H$ and an observable $\hat{O}$ such that, for all $i\in \{1,...,m\}$ and $j\in \{1,...,n\}$, letting $|\bm{x}_{i,j}^* \rangle =e^{-iHj}|\bm{x}_{i,j} \rangle $ and $c(\bm{x}_{i,j})=\langle^* \bm{x}_{i,j}|\hat{O}|\bm{x}_{i,j}^* \rangle$, we have $|c(\bm{x}_{i,j})-c(\bm{x}_{i', j'})|<\epsilon$ for all $i, i', j$ and $j'$.
\\
\\
\textbf{Question}: a. Is there a set of states $\Gamma=\{|\bm{x}_{i,j} \rangle\ : 1\leq i \leq m, 1\leq j \leq n \}$ that is not $\epsilon$-centered for some $\epsilon>0$?
\\
b. Find a natural condition that implies being $\epsilon$-centered.
\\
c. Given $\epsilon>0$ and a set $X$ of multidimensional time series, is there always a state embedding that transforms $X$ to an $\epsilon$-centered set of states?
% \begin{table}[]
% \label{tab:algocomponents}
% \begin{tabular}{p{0.1\linewidth}  p{0.5\linewidth}  p{0.4\linewidth}}
% \toprule
% Algorithm component & Description \\ \midrule
% \multirow{2}{*}{$W(\boldsymbol{\alpha})$} &
%   \multirow{2}{*}{Parameterized unitary used in the rewinding operators where $\boldsymbol{\alpha}$ is a vector of parameters.} & \\
%                     &             &                                 \\
%                     &             &                                 \\
% $D(\boldsymbol{\epsilon}, t_j)$ &
%   Diagonal unitary encoded with time-point $t_j$ and generated with the eigenvalues $\boldsymbol{\epsilon}$ of a diagonal non-unitary matrix $M$ . &
%   $k$-local approximation from Ref {[}{]}. \\
%                     &             &                                 \\
% $\hat{O}_{\boldsymbol{\eta}}$ &
%   Observable with parameters vector $\boldsymbol{\eta} = [\eta_1, \eta_2, \ldots, \eta_p]$. &
%   $\eta_1 - \frac{1}{n} \sum_{i=1}^{n} \sigma_z^i$ \\
%                     &             &                                 \\
% $U[\boldsymbol{x}_i(t_j)]$ &
%   Unitary used to embed a single point of a classical time series $\boldsymbol{x}_i(t_j)$ to generate the state $| boldsymbol{x}_i$(t\_j) \textbackslash{}rangle\$. &
%   $R_y[\boldsymbol{x}_i(t)]$ encoding. \\ \bottomrule
% \end{tabular}
% \caption{The various components of QVR and their descriptions.}
% \end{table}

\section{Univariate time series: numerical experiments \label{sec:univariate}}

The first use case in the main text is anomaly detection in synthetic univariate time series. The purpose of the case was to provide a didactic demonstration of how anomaly scores are calculated for simple time series. In this Section, we provide relevant studies on the effect of changing different parameters of the algorithm including: (i) the number of terms $N_E$ used in the calculation of the cost function (ii) the effect of varying $\boldsymbol{\mu}$ and $\boldsymbol{\sigma}$ on the cost function and (iii) the performance of different classical optimizers in the training process. \par

All the calculations in this section were obtained using an ideal quantum circuit simulator from \texttt{Pennylane} \cite{Bergolm2018}: \texttt{lightning.qubit}. With this simulator, we perform exact statevector algebra (with no need for shots) to obtain exact quantum expectation values. The penalty function hyperparamerer $\bm{\tau}$ is set to 5 for all simulations.

\subsection{Converging the classical expectation \label{subsec:univariate_expec}}

\begin{figure}[h]
    \centering
    \includegraphics[width=\linewidth]{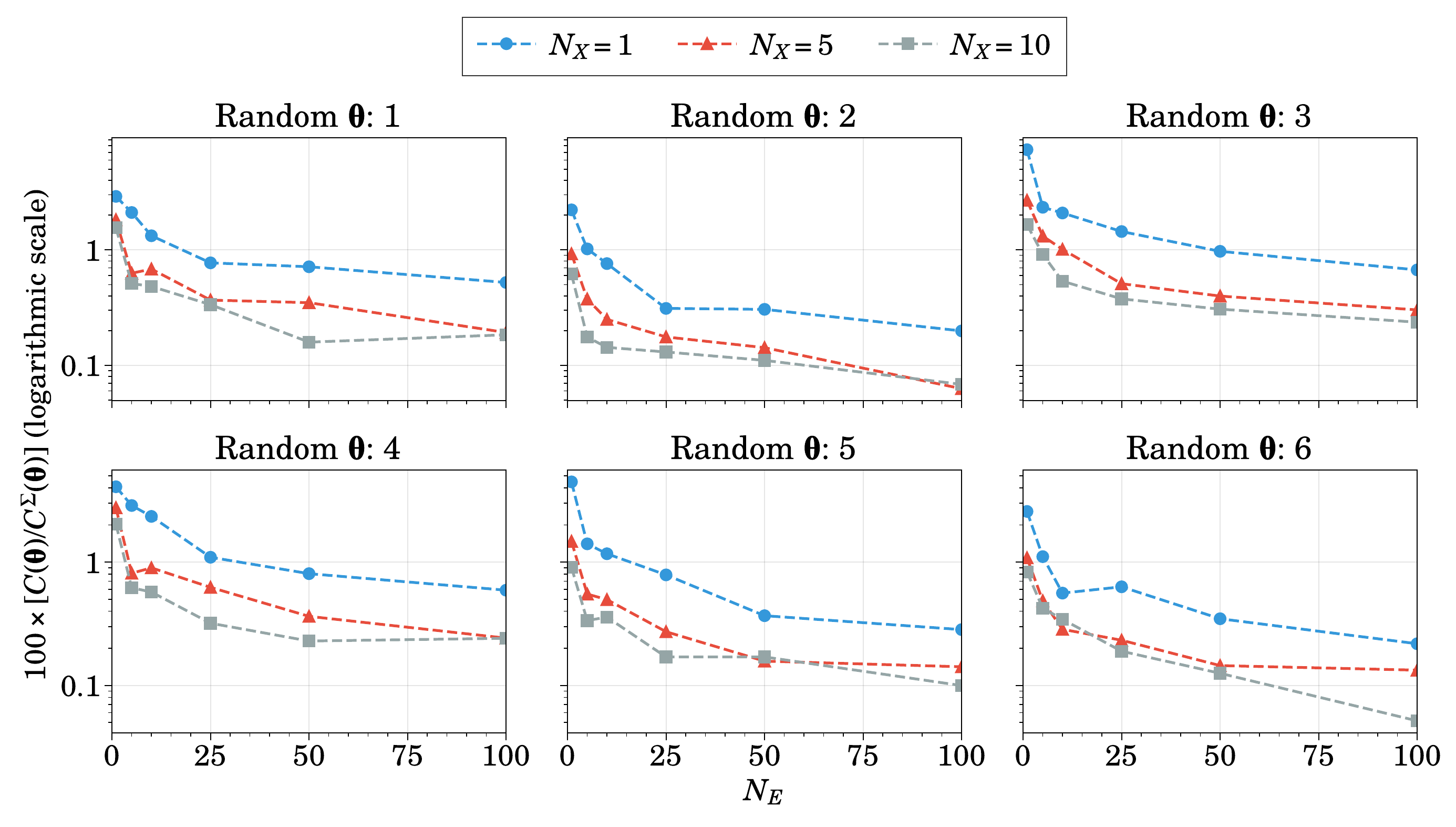}
    \caption{Convergence of the stochastic error in the cost function (logarithmic scale) with increasing the number of terms in the classical expectation estimation for univariate synthetic training data. The error is measured as the percentage ratio of the cost function to the standard deviation - $ C^{\Sigma}(\bm{\theta})$ - measured over 100 repeats. The same experiment is run for six random initializations of the model parameters and three values of the time series batching parameter: $N_X \in \{1, 5, 10 \}$. The time point batch parameter is fixed at $N_{\mathcal{T}} = 10$. }
    \label{fig:cost_convergence}
\end{figure}

Figure \ref{fig:cost_convergence} shows the convergence of the cost function error (as given by the measured standard deviation over 100 repeats) as a function of the number of terms used to compute the classical expectation value for three different batch approximations and for six random sets of model parameters $\bm{\theta}$. As expected, the error decreases with increasing $N_X$ and $N_E$. This effect is also observed with increasing $N_{\mathcal{T}}$ but is not shown here. To strike a balance between accuracy and computational load, we choose $N_X = 5$, $N_{\mathcal{T}} = 10$ and $N_E = 5$. This leads to an error $\sim$ 1\% for most choices of $\bm{\theta}$. \par

\subsection{The role of $\mathbf{\mu}$ and $\mathbf{\sigma}$ \label{subsec:univariate_mu_sigma}}

\begin{figure}[h]
    \centering
    \includegraphics[width=\linewidth]{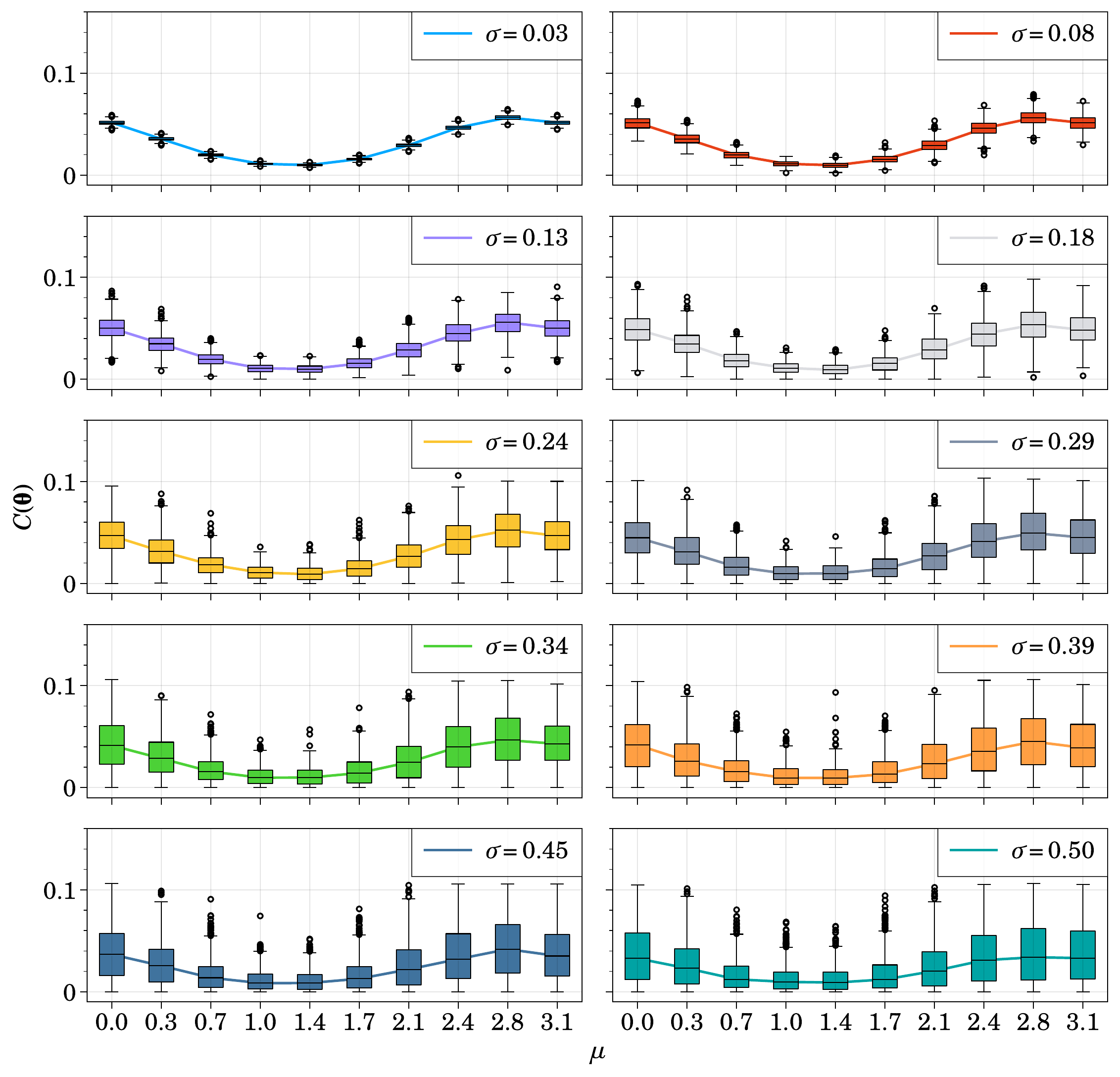}
    \caption{Evolution of the cost function with $\mu$ for different values of $\sigma$ using univariate synthetic training data at constant $\boldsymbol{\alpha}$.}
    \label{fig:expval_mu_sigma}
\end{figure}

Using the $N_E$, $N_{\mathcal{T}}$ and $N_X$ described above, Fig. \ref{fig:expval_mu_sigma} shows how the cost function varies with different $\bm{\mu}$ and $\bm{\sigma}$ when \textit{all other} other parameters in $\bm{\theta}$ are held constant. In this case, $\bm{\mu}$ $\bm{\sigma}$ are chosen to be scalar (i.e, $\bm{\epsilon}$ is drawn from a single normal distribution). Each point is the average of 100 repeats and the spread is given as a box plot. It can be seen that the cost function is smooth function of $\bm{\mu}$. Intuitively, at a fixed $\bm{\alpha}$, the spread is an increasing function of $\bm{\sigma}$; as the distribution of rewinding operators expands, a higher $N_E$ will be needed to compute the classical expectation to the same level of error. \par

\subsection{Comparing derivative-free optimizers \label{subsec:univariate_derivative_free}}

\begin{figure}[h]
    \centering
    \includegraphics[width=\linewidth]{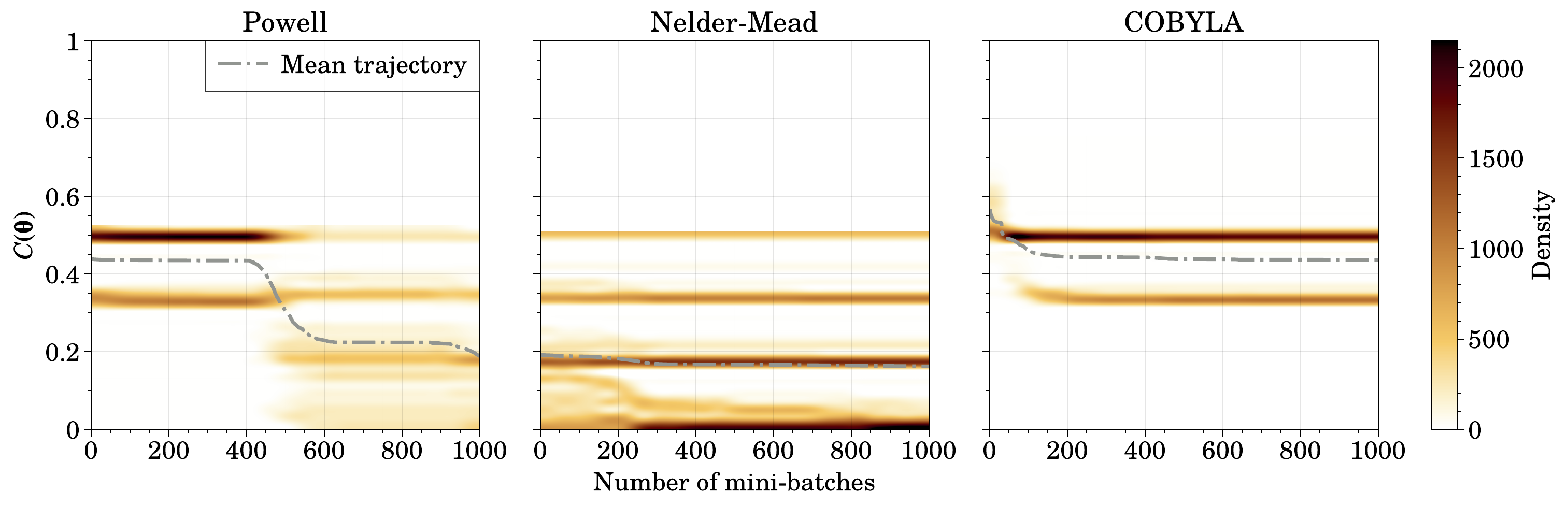}
    \caption{$C(\bm{\theta})$ as a function of the number of mini-batches for three derivative-free classical optimizers. The density heat map is the result of 100 separate training events each with randomized initial model parameters.}
    \label{fig:training_benchmark_univariate}
\end{figure}

Using the $N_E$, $N_{\mathcal{T}}$ and $N_X$ described above, Fig. \ref{fig:training_benchmark_univariate} shows the performance of different derivative-free classical optimizers used in training for the didactic univariate problem. Using Powell \cite{Powell1964}, Nedler-Mead \cite{Nelder1965} and COBYLA \cite{Powell1994}, each optimizer is used in training 100 times with different random initial $\bm{\theta}$. Each run terminates after 1000 mini-batch iterations.

As can be seen from Fig. \ref{fig:training_benchmark_univariate}, the performance of the three optimizers are very different. Notably, the density color map indicates the propensity for optimization to stall. This stalling is most common for Nelder-Mead and COBYLA. Although Nelder-Mead does achieve on average (see the mean trajectory lines on Fig. \ref{fig:training_benchmark_univariate}) lowest costs, Powell shows the most systematic decreases an the number of mini-batches increases.

\subsection{Learning noisy sinusoids \label{subsec:univariate_noisy_sinusoid}}
Here we present another study on univariate time series, examining the case of learning noisy sinusoids and testing on different sinusoids. Our training data consists of vectors $X = \{x_i : i=1,...,100\}$ of the form $x_i=(\sin(0), \sin(\frac{2\pi}{50}), \sin(\frac{4 \pi}{50}),...,\sin(2\pi)+(\epsilon_i,...,\epsilon_i)$ where $\epsilon_i \sim \mathcal{N}(0.1, 0.1)$. Using $N_X = 5$, $N_{\mathcal{T}} = 10$, $N_E = 10$ and $\bm{\tau} = 20$, the model is trained for 200 mini-batches using the Powell optimizer. Classical data is mapped to quantum states using $R_y[x_i(t_j)] \otimes I$ embedding on 2 qubits and $W(\bm{\alpha})$ is implemented with a quantum circuit consisting three layers of the Ans\"{a}tz suggested in \cite{Schuld2020}. \par  

We have three testing sets of time series:
\begin{enumerate}
    \item  $R = \{r_i : i=1,...,50\}$ where each $r_i$ is defined as the $x_i$ in the training set, but are not identical to the training set since $\epsilon_i$ are random.
    
    \item $W = \{w_i : i=1,...,50\}$ where each $w_i$ has the form $w_i=(\cos(0), \cos(\frac{2\pi}{50}), \cos(\frac{4 \pi}{50}),...,\cos(2\pi)+(\zeta_i,...,\zeta_i)$ where $\zeta_i \sim \mathcal{N}(0.1, 0.1)$.
    
    \item $Z = \{z_i : i=1,...,50\}$ where each $z_i$ has the form $z_i=(\sin(0), \sin(\frac{2\pi}{50}), \sin(\frac{4 \pi}{50}),...,\sin(2\pi)+(\xi_i,...,\xi_i)$ where $\xi_i \sim \mathcal{N}(0.1, 0.5)$.
    
\end{enumerate}

\begin{figure}[h]
    \centering
    \includegraphics[width=\linewidth]{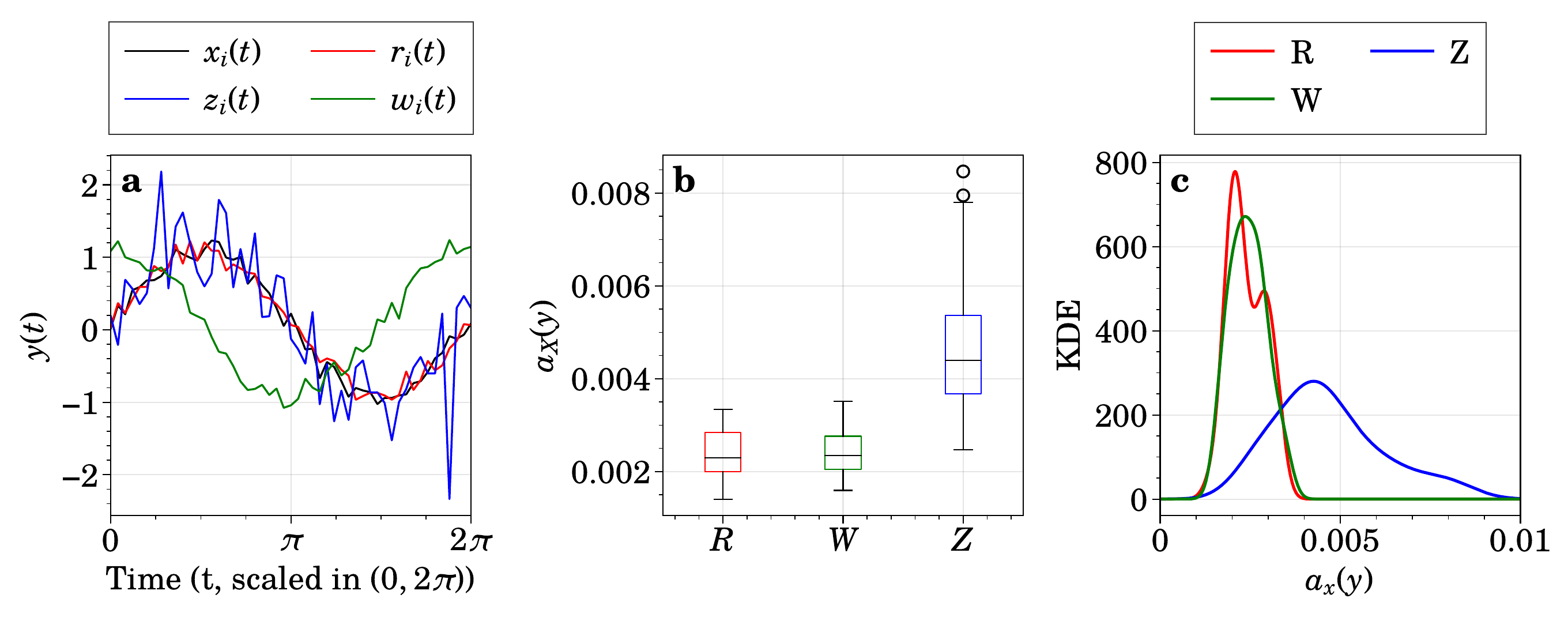}
    \caption{Experiments with noisy univariate sinusoids. (a) an example time series from each of the sets $X$, $R$, $W$ and $Z$. (b) box plots of the anomaly scores obtained from each testing set. (c) The Kernel Density Estimate (KDE) on the distribution of anomaly scores for each of the testing sets.}
    \label{fig:noisy_sinusoids}
\end{figure}

An example of each of the above time series is shown in Fig. \ref{fig:noisy_sinusoids}(a). As depicted in Fig. \ref{fig:noisy_sinusoids}(b) and \ref{fig:noisy_sinusoids}(c), testing sets $R$ and $W$ have very similar anomaly score distribtutions. On average, time series in $R$ and $W$ have lower anomaly scores than $Z$. With this information, we can infer what the notion of ``anomalous" was learnt in training. That is, despite being largely dissimilar in terms of euclidean distance,  series in $R$ and $W$ are have similar anomaly scores. However, the larger noise profile present in $Z$ produces large anomaly scores. It is clear then that the learnt notion of "anomalous" is not related to a simple euclidean distance between the average of training and testing time series at each point in time.

\section{The role of the penalty function $P_{\mathbf{\tau}}(\mathbf{\sigma})$ \label{sec:multivariate_synthetic_static}}

\subsection{The interaction of $P_{\mathbf{\tau}}(\mathbf{\sigma})$ with latent spaces \label{subsec:latent}}

We note that our algorithm relies on an idea somewhat analogous to a latent space, on which we shall now elaborate. For every $\boldsymbol{\alpha}$, our construction induces a mapping $R^b \rightarrow U(b)$ whose range we shall later denote $Im(f_{\bm{\alpha}})$. Given the optimized parameters $\boldsymbol{\alpha}^*$ and a time series $\bm{x}_i \in X$, each $|\bm{x}_i(t) \rangle$ is evolved using an appropriate unitary from $Im(f_{\bm{\alpha}})$. The guiding idea is that, given the optimized parameters, once we calculate the $\Omega^2(\bm{x}_i(t), \boldsymbol{\alpha}^*, \boldsymbol{\epsilon}^*, \boldsymbol{\eta}^*)$, then they should be close to each other. Below we suggest a heuristic justification for this.
\\
\\
For a fixed $\alpha$, consider the mapping $f_{\alpha}: R^b \rightarrow U(b)$ defined by \begin{equation}f_{\alpha}(\bm \epsilon):=W(\alpha)^{\dagger}e^{-iM(\bm \epsilon)}W(\alpha) \end{equation} where $M(\bm \epsilon)$ is a diagonal matrix whose diagonal entries are the elements of $\bm \epsilon$. Observe that $f_{\alpha}$ is a continuous group homomorphism from $(R^n, +)$ to $U(n)$: $f_{\alpha}(\bm \epsilon+ \bm{\epsilon'})=W(\alpha)^{\dagger}e^{-iM(\bm \epsilon + \bm{\epsilon'})}W(\alpha)=W(\alpha)^{\dagger}e^{-iM(\bm \epsilon)} e^{-iM(\bm{\epsilon'})}W(\alpha)=f_{\alpha}(\bm \epsilon) f_{\alpha}(\bm{\epsilon'})$.
\\
\\
Note also that $f_{\alpha}(\bm \epsilon)=I$ iff $e^{-i\epsilon_i}=1$ for all $i$: multiplying by $W(\alpha)$ on the left and $W(\alpha)^{\dagger}$ on the right, $f_{\alpha}(\bm \epsilon)=I$ iff $e^{-iM(\bm \epsilon)}=I$. As $-iM(\bm \epsilon)$ is diagonal, $e^{-iM(\bm \epsilon)}$ is a diagonal matrix whose diagonal terms have the form $e^{-i \epsilon_i}$, and so $f_{\alpha}(\bm \epsilon)=I$ iff $e^{-i \epsilon_i}=1$ for all $i$. Recall that $e^z=1$ exactly when $z=2 \pi i k$ for some $k\in Z$. Therefore, the kernel of $f_{\alpha}$ is the lattice $2 \pi Z^n$ (so in particular, $f_{\alpha}(\bm \epsilon)=f_{\alpha}(\bm{\epsilon'})$ iff $\bm \epsilon - \bm{\epsilon'} \in 2\pi Z^n$) and $(R^n, +)/2\pi Z^n$ is isomorphic as a group to $Im(f_{\alpha})$.
\\
\\
We shall refer to $Im(f_{\alpha})$ as our \textbf{latent space of unitary matrices}. Given $\bm{\sigma}$ and $\bm{\mu}$ as in the formalism, let $\mathcal{N}^*(\bm \sigma, \bm \mu)$ be the distribution on $R^b$ obtained from all the $\mathcal{N}(\sigma_i, \mu_i)$. For a measurable set $X\subseteq R^b$, denote $\mu_{\mathcal{N}^*(\bm \sigma, \bm \mu)}(X):=\int_X f dm$ where $f$ is the density function of $\mathcal{N}^*(\bm \sigma, \bm \mu)$ and $m$ is the Lebesgue measure. Consider the $\sigma$-algebra $\mathcal A$ of subsets $X\subseteq Im(f_{\alpha})$ such that $\mu_{\mathcal{N}^*(\bm \sigma, \bm \mu)}(f_{\alpha}^{-1}(X))$ is well-defined, then we can define a probability measure on the latent space $Im(f_{\alpha})$ by letting $\mu(X):=\mu_{\mathcal{N}^*(\bm \sigma, \bm \mu)}\{ \bm \epsilon \in R^b : f_{\alpha}(\bm \epsilon) \in X\}$ for a subset $X \subseteq Im(f_{\alpha})$ from $\mathcal A$. Therefore, by considering different values of $\bm \sigma$ and $\bm \mu$, we consider different probability distributions on the latent space of unitary matrices. 
\\
\\
Fix a maximal subset $S\subseteq R^b$ of $2\pi Z$-inequivalent vectors and fix an interval $I \subseteq R$. For a time series $\bm x$, we say that $\bm \epsilon \in R^b$ is $(I, \bm x)$-good if $C_2(\bm x, \alpha, \bm \epsilon)^- \in I$.
\\
\\
We formulate below three conjectures:
\medskip
\\
\textbf{Conjectures}: Let $I$ and $S$ be as above.
\begin{enumerate}
    \item There exists some $\delta<<1$ such that, for every time series $\bm x$, the Lebesgue measure of the $(I, \bm x)$-good vectors in $S$ is $<\delta$.
    
    \item There exists some $\delta<< m(S)$ such that, for every two times series $\bm x$ and $\bm y$, the difference between the Lebesgue measures of the $(I, \bm x)$-good vectors and $(I, \bm y)$-good vectors is at most $\delta$.
    
    \item For every two time series $\bm x$ and $\bm y$, the sets of $(I, \bm x)$-good vectors and $(I, \bm y)$-good vectors have the same Lebesgue measure.
\end{enumerate}
\medskip
Note that both $(3)$ and $(1)$ imply $(2)$, and note also that the conjectures do not depend on the choice of $S$. Now each of the above three conjectures implies a variant of the following heuristic: By minimizing $\sigma$, the $\bm{\epsilon}$'s that contribute greatly to $C_2(\bm x, \alpha)$ are less likely to be $(I, \bm y)$-good for $\bm y \neq \bm x$ and a small interval $I$ that contains $C_2(\bm x, \alpha)$, thus increasing $\bm y$'s anomaly score.

\subsection{Tuning the hyperparameter of $P_{\mathbf{\tau}}(\mathbf{\sigma})$ in a static anomaly detection variation \label{subsec:static}}

Although QVR is intended to deal with time series data, it can be trivialized to a static anomaly detection algorithm by simply considering ``time series" with only one time point $x_i(t_j), j=0$ where the value of $t_j$ is arbitrary (we choose $t_j = 1$ for the forthcoming).

\begin{figure}
    \centering
    \includegraphics[width=\linewidth]{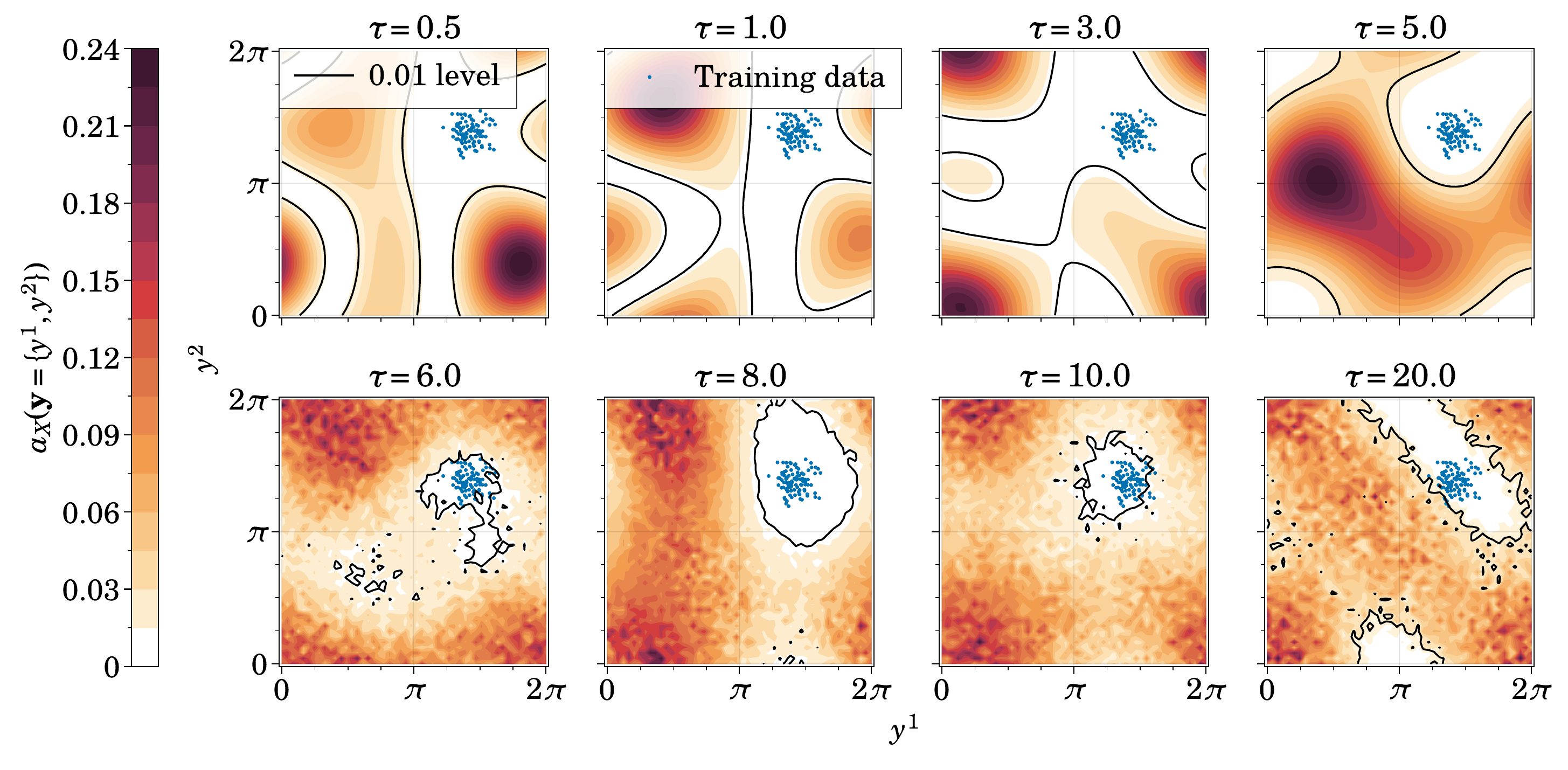}
    \caption{The optimal cost landscapes of static (i.e at a single time point, $t = 1$) 2D feature spaces over the input ranges of the features $y^1$ and $y^2$ for different values of the penalty function hyperparameter $\tau$. The feature range is $2\pi$-periodic because $R_x(\theta)$ embedding is used.}
    \label{fig:2d_feature_static}
\end{figure}

This static anomaly detection variation is useful for understanding the effects increasing the single penalty hyperparameter $\tau$. An investigation into this is shown in Fig. \ref{fig:2d_feature_static}. Training data is chosen to be two-dimensional ``blobs" of 100 points (see blue points on Fig. \ref{fig:2d_feature_static}) generated using \texttt{scikit-learn} \cite{Pedregosa2012} with a blob standard deviation of $\pi/4$ centered close to the coordinates ($3\pi/2$, $3\pi/2$). Data embedding and the implementation of $W(\bm{\alpha})$ are the same as noisy sinusoid univariate case. Cycling through eight values of $\tau$: $(0.5, 1.0, 3.0, 5.0, 6.0, 8.0, 10.0, 20.0)$, the model is trained for 200 mini-batches using the Powell optimizer setting $N_X = 10$, $N_E = 10$ and (trivially) $N_{\mathcal{T}} = 1$.

As can be seen from Fig. \ref{fig:2d_feature_static}, the effect of increasing $\tau$ is broadly to increasingly surround the training data with a tightening ``lasso" defined by the $a_X(\bm{y}) = 0.01$ level. Consistent with the discussion in Section \ref{subsec:latent}, we see that data outside of the training set is more likely to be anomalous. Interestingly, we also see higher frequency variations in the 2D surface defined by $a_X(\bm{y}=\{y^1, y^2 \})$ as $\tau$ is increased.

\section{Multivariate Cryptocurrency models \label{sec:realMultivariate}}

\subsection{The data set and its preparation \label{subsec:cryptodataset}}

The data used in the crpytocurrency use case is a combination of the Bitcoin (BTC) vs. Tether (USDT) time series trade data with the data of large transactions on the BTC and USDT blockchains. The trade data comes from the ``Binance Full History" Kaggle dataset: (\url{https://www.kaggle.com/datasets/jorijnsmit/binance-full-history}), which obtains its data from the official API of the Binance cryptocurreny exchange (\url{https://github.com/binance/binance-spot-api-docs/blob/master/rest-api.md}). Meanwhile, the large blockchain transactions data comes from the Whale Alert Twitter account (\url{https://twitter.com/whale_alert}), which obtains its data from various public cryptocurrency blockchains.

From the Binance Full History Data, we only extracted the BTC-USDT time series for the time period between 2019-06-06 and 2021-11-24. Moreover, we only extracted the following features: open, high, low, close, volume, number\_of\_trades, taker\_buy\_base\_asset\_volume. The frequency of the time series is 1 min. 

The data of large transactions on blockchains is extracted from the public blockchain ledgers and continually posted on Whale Alert twitter account (which can be accessed via Twitter API or Whale Alert API). From this data, we only keep tweets which mention \#Binance and one of the two coins of interest: Bitcoin (\#BTC) and Tether (\#USDT). From within these tweets, we extract the following variables: the amount of BTC or USDT coins transferred, the worth of these coins in USD at the time of the transaction, and the time at which the transaction occurred.

In combining these datasets, the timestamps of large cryptocurrency transactions are used to extract 3-hour chunks of data from the BTC-USDT timeseries. In particular, for a given transaction, one hour interval before the timestamp and 2 hours after the timestamp are extracted. The rationale for extracting an hour before the transaction is that not only deposits, but also withdrawals of cryptocurrency are tracked. A large withdrawal could be preceded by unusual market activity on the exchange.

The dataset created this way can be contaminated in the sense that more than one large cryptocurrency transaction could have occurred within a given 3-hour time change or sometime before it. Thus, to generate an uncontaminated subset of the data, we require that there be only a single large cryptocurrency transaction within 3 hours before the transaction and 2 hours after it. 

Of the extracted features, in main article, we make use of at most three. These three features are:

\begin{enumerate}
    \item \textbf{Mean deviation of the open price}: The mean value of the opening BTC-USDT price is taken over the time window. Then, each point in the time series is the difference with the mean value.
    
    \item \textbf{Cumulative volume}: A cumulative sum of the trading volume over the time window.
    
    \item \textbf{Cumulative taker by base asset (TBBA) volume} A cumulative sum of the TBBA volume over the time window.
    
\end{enumerate}

For many choices of embedding unitary $U[\bm{x}_i(t)]$ acting on some state $|\psi \rangle$, the expectation value defined by $\langle \psi|\hat{O}_{\bm{\eta}}| \psi \rangle$ becomes defined $\mod 2\pi$ (indeed, this effect can be seen in Fig. \ref{fig:2d_feature_static}). Because of this, all of the above time series need to be re-scaled in this domain. To do so, we use a ``min-max-scaling" procedure. That is, at each time point $t_j$ the maximum value of $x_i^d(t_j) \forall i$ ($x_{i, \text{max}}^d(t_j)$) is assigned a value of $\pi$ and the minimum value ($x_{i, \text{min}}^d(t_j)$) is assigned a value of $-\pi$. All other values at this $t_j$ are now scaled between these two limits, i.e

\begin{equation}
    \text{rescaled}[x_i^d(t_j)] = 2\pi \frac{x_i^d(t_j) - x_{i, \text{min}}^d(t_j)}{x_{i, \text{max}}^d(t_j) - x_{i, \text{min}}^d(t_j)} - \pi
\end{equation}

The model using all three of the re-scaled time series is referred to as a trivariate model and the bivariate model uses only the mean deviation of the open price and cumulative volume.

All of the above re-scaled time series can be found in the \texttt{data} directory at \url{https://github.com/AgnostiqHQ/QuantumVariationalRewinding} or hosted on Zenodo (\url{https://doi.org/10.5281/zenodo.7258627}) 

\subsection{Training the bivariate and trivariate models \label{subsec:training_crypto}}

\begin{figure}
    \centering
    \includegraphics[width=\linewidth]{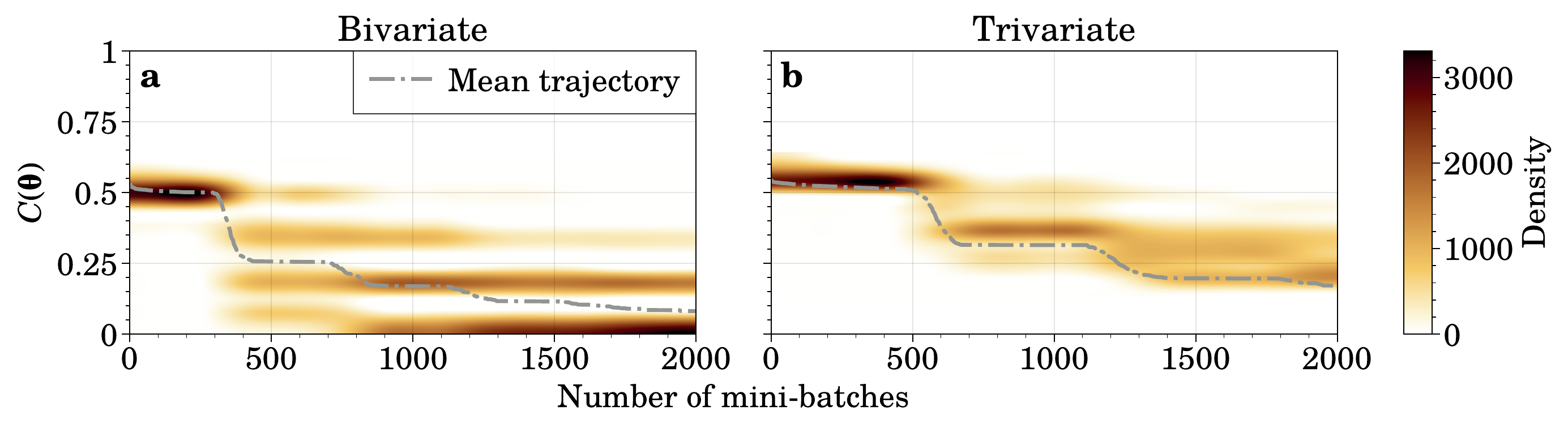}
    \caption{Training the cryptocurrency time-series models with the Powell optimizer. (a) The bivariate model. (b) The trivariate model}
    \label{fig:training_bi_tri}
\end{figure}

Using the time series defined above, each model (bivariate and trivariate) is trained 50 times with different random initial $\bm{\theta}$ for 2000 mini-batches. We use $N_E = N_{\mathcal{T}} = N_{\mathcal{X}} = 10$, $\tau = 5$, $R_y[x_i^1(t_j)] \otimes R_y[x_i^2(t_j)] \otimes \ldots R_y[x_i^d(t_j)]$ encoding and three layers of the Ans\"{a}tz proposed in \cite{Schuld2020}. The bivariate model uses 2 qubits while the trivariate models requires 3 qubits. $C(\bm{\theta})$ as a function of mini-batch iterations for each model is shown in Fig. \ref{fig:training_bi_tri}. Broadly, we see that the bivariate model is able to reach lower costs that the trivariate model within the number of batch iterations. We hypothesize that this is because there are a greater number of total parameters for the trivariate model. Interestingly, the training the bivariate model does stall more often than the trivariate case, which, on average shows a comparatively smooth decreasing function of $C(\bm{\theta})$ with mini-batch iterations.

\subsection{Details of superconducting transmon quantum computers and Qiskit Runtime \label{subsec:transmon}}

In the main article, we perform experiments using IBM’s \texttt{Falcon r5.11H} family of superconducting transmon QPUs. Below in Table \ref{tab:QPUstats}, we display their properties. Note that these figures were taken at the time we began the first runs (7th July 2022) but will vary heavily over time according to the calibration state of the machine and other factors.

\begin{table}[h]
\centering
\begin{tabular}{@{}cccccc@{}}
\toprule
Quantum Processing Unit & Quantum Volume & Speed [CLOPS] & Qubits & Mean CNOT error      & Mean $T_2$ coherence [$\mu$s] \\ \midrule
\texttt{ibm\_lagos}     & 32             & 2700          & 7      & $7.254\times10^{-3}$ & 92.96                         \\
\texttt{ibm\_perth}     & 32             & 2900          & 7      & $1.312\times10^{-2}$ & 124.58                        \\
\texttt{ibmq\_jakarta}  & 16             & 2400          & 7      & $8.370\times10^{-3}$ & 43.98                         \\ \bottomrule
\end{tabular}
\caption{Some salient properties of the superconducting transmon quantum computers used in the main article.}
\label{tab:QPUstats}
\end{table}

The use these QPU's effectively (i.e maximizing speed), we implemented QVR with a custom \texttt{Qiskit Runtime} program. Such programs are deployed to a CPU ``Near" the QPU such that communication latency between the two is much reduced (without this, data has to travel from a user's local machine to a remote QPU via the internet, which incurs a large latency). To implement such a program, we created a conversion layer from our \texttt{Pennylane} implementation to \texttt{Qiskit} style quantum circuits. These circuits are then used in the custom \texttt{Qiskit Runtime} program. We also use error mitigation in the main article which makes use of code in the \texttt{Qiskit-research} repository \url{https://github.com/qiskit-research/qiskit-research}.

\begin{@fileswfalse}
\bibliography{ms}
\end{@fileswfalse}

\end{document}